\documentclass[10pt, twocolumn]{IEEEtran}
\usepackage[pdftex,final]{graphicx}
\usepackage{amsmath}
\usepackage{amsthm}
\usepackage{lscape}
\usepackage{latexsym}
\usepackage{amssymb}
\usepackage{bm}
\usepackage{epstopdf}
\usepackage{bbm}
\usepackage{cite}
\usepackage{url}
\usepackage{color}
\usepackage[caption=false,font=footnotesize]{subfig}
\usepackage{lmodern}
\usepackage{tikz}
\usetikzlibrary{shapes.geometric, arrows, positioning, fit}
\newcommand\blfootnote[1]{%
	\begingroup
	\renewcommand\thefootnote{}\footnote{#1}%
	\addtocounter{footnote}{-1}%
	\endgroup
}

\theoremstyle{definition}

\newcommand{\argmin}{\operatornamewithlimits{arg\,min}}

\title{Deep Joint Source-Channel Coding for \\ Wireless Image Transmission}

\author{\IEEEauthorblockN{Eirina Bourtsoulatze, David Burth Kurka and Deniz G\"und\"uz} }

\begin{document}
\maketitle

\vspace{-0.7cm}
\begin{abstract}
We propose a joint source and channel coding (JSCC) technique for wireless image transmission that does not rely on explicit codes for either compression or error correction; instead, it directly maps the image pixel values to the complex-valued channel input symbols. We parameterize the encoder and decoder functions by two convolutional neural networks (CNNs), which are trained jointly, and can be considered as an \textit{autoencoder} with a non-trainable layer in the middle that represents the noisy communication channel. Our results show that the proposed deep JSCC scheme outperforms digital transmission concatenating JPEG or JPEG2000 compression with a capacity achieving channel code at low signal-to-noise ratio (SNR) and channel bandwidth values in the presence of additive white Gaussian noise (AWGN). More strikingly, deep JSCC does not suffer from the ``cliff effect'', and it provides a graceful performance degradation as the channel SNR varies with respect to the SNR value assumed during training. In the case of a slow Rayleigh fading channel, deep JSCC learns noise resilient coded representations and significantly outperforms separation-based digital communication at all SNR and channel bandwidth values.

\end{abstract}


\bstctlcite{ref_papers:BSTcontrol}
\blfootnote{E. Bourtsoulatze is with the Communications and Information Systems Group, Department of Electronic and Electrical Engineering, University College London, London, UK. D. Burth Kurka and D. G\"und\"uz are with the Information Processing and Communications Laboratory, Department of Electrical and Electronic Engineering, Imperial College London, London, UK. Part of this work was done while the first author was with the Information Processing and Communications Laboratory, Imperial College London.}
\blfootnote{E-mails: {\tt e.bourtsoulatze@ucl.ac.uk}, {\tt  d.kurka@imperial.ac.uk, d.gunduz@imperial.ac.uk}}
\blfootnote{This work has been funded by the European Union's Horizon 2020 research and innovation programme under the Marie Sk\l{}odowska-Curie fellowship (grant agreement No. 750254) and by the European Research Council (ERC) through the Starting Grant BEACON (grant agreement No. 677854).}

\section{Introduction}
\label{sec:introduction}

Modern communication systems employ a two step encoding process for the transmission of image/video data (see Fig. \ref{fig:comm_system1} for an illustration): \textit{(i)} the image/video data is first compressed with a source coding algorithm in order to get rid of the inherent redundancy, and to reduce the amount of transferred information; and \textit{(ii)} the compressed bitstream is first encoded with an error correcting code, which enables resilient transmission against errors, and then modulated. Shannon's \textit{separation theorem} proves that this two-step source and channel coding approach is optimal theoretically in the asymptotic limit of infinitely long source and channel blocks \cite{Cover:book}. While in practical applications joint source and channel coding (JSCC) is known to outperform the separate approach \cite{jscc:handbook}, separate architecture is attractive for practical communication systems thanks to the modularity it provides. Moreover, highly efficient compression algorithms (e.g. JPEG, JPEG2000, WebP \cite{Google:WebP}) and near-optimal channel codes (e.g. LDPC, Turbo codes) are employed in practice to approach the theoretical limits.  However, many emerging applications from the Internet-of-things to autonomous driving and to tactile Internet require transmission of image/video data under extreme latency, bandwidth and/or energy constraints, which preclude computationally demanding long-blocklength source and channel coding techniques.

We propose a JSCC technique for wireless image transmission that directly maps the image pixel values to the complex-valued channel input symbols. Inspired by the success of unsupervised deep learning (DL) methods, in particular, the autoencoder architectures \cite{BengioFTML, GoodfellowDL2016}, we design an end-to-end communication system, where the encoding and decoding functions are parameterized by two convolutional neural networks (CNNs) and the communication channel is incorporated in the neural network (NN) architecture as a non-trainable layer; hence, the name \textit{deep JSCC}. Two channel models, the additive white Gaussian noise (AWGN) channel and the slow Rayleigh fading channel, are considered in this work due to their widespread adoption in representing realistic channel conditions. The proposed solution is readily extendable to other channel models, as long as they can be represented as a non-trainable NN layer with a differentiable transfer function.

DL-based methods, and, particularly, autoencoders, have recently shown remarkable results in image compression, achieving or even surpassing the performance of state-of-the-art lossy compression algorithms. Ball\'{e} \textit{et al.} \cite{Balle:ICLR:17} propose an end-to-end optimized image compression method, consisting of a nonlinear analysis transformation, a uniform quantizer, and a nonlinear synthesis transformation. Their method exhibits better rate-distortion performance than JPEG and JPEG2000 in most images, while the visual quality, as captured by the MS-SSIM metric, improves for all test images and over all bitrate values. A compressive autoencoder is used in \cite{TheisICLR2017}, where the authors propose to use a proxy of the quantization step only in the backward propagation, while keeping the rounding in the forward step.  The authors of  \cite{RippelICML2017} complement the autoencoder based compression architecture with adversarial loss to achieve realistic reconstructions and improve the visual quality.  Cheng \textit{et al.} \cite{ChengPCS2018} present a convolutional autoencoder  based lossy image compression architecture, which achieves on average a 13.5\% rate saving versus JPEG2000 on the Kodak image dataset. The advantage of DL-based methods for lossy compression versus conventional compression algorithms lies in their ability to extract complex features from the training data thanks to their deep architecture, and the fact that their model parameters can be trained efficiently on large datasets through backpropagation. While common compression algorithms, such as JPEG, apply the same processing pipeline to all types of images (e.g., DCT transform, quantization and entropy coding in JPEG), the DL-based image compression algorithms learn the statistical characteristics from a large training dataset, and optimize the compression algorithm accordingly, without explicitly specifying a transform or a code.

At the same time, the potential of DL has also been capitalized by researchers to design novel and efficient coding and modulation techniques in communications. In particular, the similarities between the autoencoder architecture and the digital communication systems have motivated significant research efforts in the direction of modelling end-to-end communication systems using the autoencoder architecture \cite{OShea:ISSPIT:16,deep:PHY}. Some examples of such designs include decoder design for existing channel codes \cite{Kim:ICLR:18, Nachmani:JSTSP:18}, blind channel equalization \cite{vae:bce}, learning physical layer signal representation for SISO \cite{deep:PHY} and MIMO \cite{deep:MIMO} systems, OFDM systems \cite{Felix:SPAWC:18,JuangWCL2018}, JSCC of text messages \cite{FarsadICASSP2018} and JSCC for MNIST images for analog storage \cite{ZarconeDCC2018}.

In this work, we leverage the recent success of DL methods in image compression and communication system design to propose a novel JSCC algorithm for image transmission over wireless communication channels. We consider both time-invariant and fading AWGN channels, and compare the performance of our algorithm to the state-of-the-art compression algorithms (JPEG and JPEG2000, in particular) combined with capacity-achieving channel codes. We show through experiments that our solution achieves superior performance in low signal-to-noise ratio (SNR) regimes and for limited channel bandwidth, over a time-invariant AWGN channel, even though the separation scheme is assumed to be operating at the channel capacity despite the short blocklengths. While we have mainly focused on the peak signal-to-noise ratio ($\mathrm{PSNR}$) as the performance measure, we show that the deep JSCC can provide even better results when measured in terms of the structural similarity index (SSIM), which better captures the perceived visual quality of the reconstructed images.  More interestingly, we demonstrate that our approach is resilient to variations in channel conditions, and does not suffer from abrupt quality degradations, known as the ``cliff effect'' in digital communication systems: deep JSCC algorithm exhibits graceful performance degradation when the channel conditions deteriorate. This latter property is particularly attractive when broadcasting the same image to multiple receivers with different channel qualities, or when transmitting to a single receiver over an unknown fading channel. Indeed, we show that the proposed deep JSCC scheme achieves a remarkable performance over a slow Rayleigh fading channel by learning coded representations robust to channel quality fluctuations and outperforms a separation-based digital transmission scheme even at high SNR and large channel bandwidth scenarios.

This is the first time an end-to-end joint source-channel coding architecture is trained for wireless transmission of high-resolution images over AWGN and fading channels. This architecture allows training for other performance measures or other source signals (e.g., video) as well. Moreover, while the training of the deep JSCC algorithm can be fairly time consuming, once the network is trained, the encoding and decoding tasks become extremely fast, compared to applying advanced image compression/decompression algorithms followed by capacity-approaching channel coding and decoding. We believe this may be key to enabling many low-latency applications that require the transmission of high data rate content at the wireless edge, such as image/video sensor data from autonomous cars or drones, or emerging AR/VR applications. We also emphasize that the employed neural network architecture is quite efficient consisting of fully convolutional layers. With the rapid advances in hardware accelerators specially optimized for CNNs \cite{AIchips, FPGA:CNN}, we believe the deep JSCC can very soon be deployed directly on mobile wireless devices.

The rest of the paper is organized as follows. In Section \ref{s:problem}, we introduce the system model, provide some background on the conventional wireless image transmission systems and their limitations, and motivate our novel approach. We introduce the proposed deep JSCC architecture in Section \ref{sec:jscc_algo}. Section \ref{s:evaluation} is dedicated to the evaluation of the performance of the proposed deep JSCC scheme, and its comparison with the conventional separate JSCC schemes over both static and fading AWGN channels. Finally, the paper is concluded in Section \ref{s:conclusions}.

\section{Background and Problem Formulation}\label{s:problem}

We consider image transmission over a point-to-point wireless communication channel. The transmitter maps the input image $\bm x\in  \mathbb{R}^n$ to a vector of complex-valued channel input symbols ${\bm z} \in \mathbb{C}^k$. Following the JSCC literature, we will call the image dimension $n$ as the \textit{source bandwidth}, and the channel dimension $k$ as the \textit{channel bandwidth}. We typically have $k < n$, which is called \textit{bandwidth compression}. We will refer to the ratio $k/n$ as \textit{bandwidth compression ratio}. Due to practical considerations in real-world communication systems, e.g., limited energy, interference, {\em etc.}, the output of the transmitter may be required to satisfy a certain power constraint, such as peak and/or average power constraints. The output signal $\bm z$ is then transmitted over the channel, which degrades the signal quality due to noise, fading, interference or other channel impairments. The corrupted output of the communication channel $\hat{\bm z }\in \mathbb{C}^k$ is fed to the receiver, which produces an approximate reconstruction $\hat{\bm x} \in \mathbb{R}^n$ of the original input image.

\begin{figure}[t]
	\begin{center}
		\subfloat[]{\resizebox {0.95\columnwidth} {!} {

 \tikzstyle{txt} = [text centered]
 \tikzstyle{box} = [rectangle, rounded corners, minimum width=2.5cm, minimum height=1.5cm, text width=2.5cm, text centered, draw=black]
 \tikzstyle{arrow} = [thick,->,>=stealth]
 \tikzstyle{fitted} = [draw=gray, thick, dotted, inner sep=0.5em]


\begin{tikzpicture}[node distance=2.3cm]

\node (x) [txt, font=\fontsize{14}{0}\selectfont] {$\bm{x}$};
\node (sencoder) [box, right of=x, font=\fontsize{12}{12}\selectfont] {Source  \\ Encoder \\ ($f_s$)};
\node (cencoder) [box, right of=sencoder, xshift=1cm,  font=\fontsize{12}{12}\selectfont] {Channel Encoder\\($f_c$)}; 
\node (modulator) [box, right of=cencoder, xshift=1cm,  font=\fontsize{12}{12}\selectfont] {Modulator\\($f_m$)};
\node (z) [txt, right of=modulator, xshift=0.5cm, font=\fontsize{14}{0}\selectfont] { };
\node (channel) [box, below of=z,  font=\fontsize{12}{12}\selectfont, yshift=0.5cm] {Noisy Channel};
\node (zhat) [txt, below of=channel, font=\fontsize{14}{0}\selectfont,yshift=0.5cm] {};
\node (demodulator) [box, left of=zhat, xshift=-0.5cm,  font=\fontsize{12}{12}\selectfont] {Demodulator\\($g_m$)};
\node (cdecoder) [box, left of=demodulator, xshift=-1cm,  font=\fontsize{12}{12}\selectfont] {Channel Decoder\\($g_c$)}; 
\node (sdecoder) [box, left of=cdecoder, xshift=-1cm,  font=\fontsize{12}{12}\selectfont] {Source Decoder\\($g_s$)};
\node (xhat) [txt, left of=sdecoder,font=\fontsize{14}{0}\selectfont] {$\bm{\hat{x}}$};
\node (tx) [fitted, fit=(sencoder) (modulator)] {};
\node at (tx.north) [above, inner sep=3mm] {Transmitter};
\node (rx) [fitted, fit=(sdecoder) (demodulator)] {};
\node at (rx.north) [above, inner sep=3mm] {Receiver};

\draw [arrow] (x) -- (sencoder);
\draw [arrow] (sencoder) --   (cencoder);
\draw [arrow] (cencoder) --  (modulator);
\draw [arrow] (modulator.east) -- node[above,font=\fontsize{14}{0}\selectfont] {$\bm{z}$} ++(1.0, 0) -| (channel);
\draw [arrow] (channel.south) -- ++(0,-1.0) -- node[below,font=\fontsize{14}{0}\selectfont] {$\hat{\bm{z}}$}(demodulator);
\draw [arrow] (demodulator) --  (cdecoder);
\draw [arrow] (cdecoder) --  (sdecoder);
\draw [arrow] (sdecoder) --(xhat);

\end{tikzpicture}

}\label{fig:comm_system1}}  \\
		\subfloat[]{%
        \resizebox {0.95\columnwidth} {!}%
        {

\tikzstyle{txt} = [text centered]
\tikzstyle{box} = [rectangle, rounded corners, minimum width=3cm, minimum height=2cm, text width=3cm, text centered, draw=black]
\tikzstyle{sbox} = [rectangle, rounded corners, minimum width=5cm, minimum height=1.5cm, text width=4.5cm, text centered, draw=black]
\tikzstyle{arrow} = [thick,->,>=stealth]


\begin{tikzpicture} [node distance=4.5cm]

\node (x) [txt, font=\fontsize{20}{0}\selectfont] {$\bm{x}$};
\node (encoder) [sbox, right =1cm of x,  font=\fontsize{19}{19}\selectfont] {Joint\\Source-Channel \\ Encoder\\($f_{\bm{\theta}}$)}; 
\node (channel) [box, right =1.5cm of encoder,  font=\fontsize{19}{19}\selectfont ] {Noisy \\ Channel \\ ($\eta$)};
\node (decoder) [sbox, right =1.5cm of channel,  font=\fontsize{19}{19}\selectfont] {Joint\\Source-Channel \\ Decoder\\($g_{\bm{\phi}}$)}; 
\node (xhat) [txt, right =1cm of decoder, font=\fontsize{20}{0}\selectfont] {$\bm{\hat{x}}$};

\draw [arrow] (x) -- (encoder);
\draw [arrow] (encoder.east)  --node[above,font=\fontsize{20}{0}\selectfont] {$\bm{z}$} (channel.west);
\draw [arrow] (channel.east) -- node[above,font=\fontsize{20}{0}\selectfont] {$\hat{\bm{z}}$}(decoder.west);
\draw [arrow] (decoder.east) -- (xhat);

\end{tikzpicture}

}%
        \label{fig:comm_system2}%
        }
	\end{center}
\caption{Block diagram of the point-to-point image transmission system: (a) components of the conventional processing pipeline and (b) components of the proposed deep JSCC algorithm.}
\label{fig:comm_system}
\end{figure}
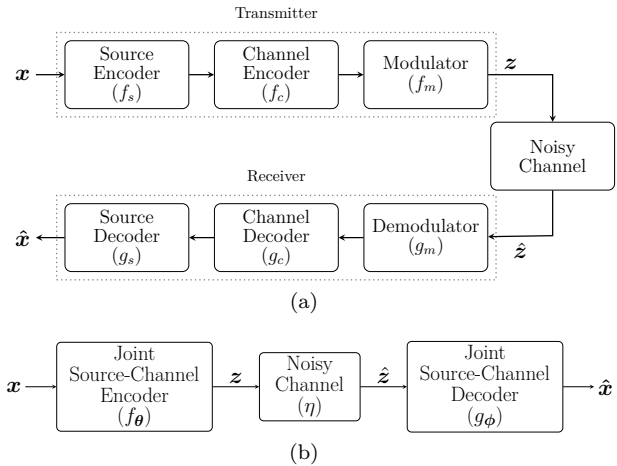

In conventional image transmission systems, depicted in Fig. \ref{fig:comm_system1}, the transmitter performs three consecutive independent steps in order to generate the signal $\bm z$ transmitted over the channel. First, the source redundancies are removed with a source encoder $f_s$, which is typically one of the commonly used compression methods (e.g., JPEG/JPEG2000, WebP). A channel code $f_c$ (e.g., LDPC, Turbo code) is then applied to the compressed bitstream in order to protect it against the impairments introduced by the communication channel. Finally, the coded bitstream is modulated with a modulation scheme $f_m$ (e.g., BPSK, 16-QAM) which maps the bits to complex-valued samples. The modulated symbols are then carried by the I and Q digital signal components over the communication link (the latter two components are often combined into a single coded-modulation step \cite{FTCIT:Fabregas}).

The decoder inverts these operations in the reverse order. It first demodulates and maps the complex-valued channel output samples to a sequence of bits (or, log likelihood ratios) with a demodulation scheme $g_m$ that matches the modulator $f_m$. It then decodes the channel code with a channel decoding algorithm $g_c$, and finally provides an approximate reconstruction of the transmitted image from the (possibly corrupted) compressed bitstream by applying the appropriate decompression algorithm, $g_s$.

Though the above encoding process is highly optimized and widely adopted in image transmission systems \cite{Thomos:TIP:06}, its performance may suffer severely when the channel conditions differ from those for which the system has been optimized. Although the source and channel codes can be designed separately, their rates are chosen jointly targeting a specific channel quality, i.e., assuming that a capacity achieving channel code can be employed, the compression rate is chosen to produce exactly the amount of data that can be reliably transmitted over the channel. However, when the experienced channel condition is worse than the one for which the code rates are chosen, the error probability increases rapidly, and the receiver cannot receive the correct channel codeword with a high probability. This leads to a failure in source decoder as well, resulting in a significant reduction in the reconstruction quality.

Similarly, the separate design cannot benefit from improved channel conditions either; that is, once the source and channel coding rates are fixed, no matter how good the channel is, the reconstruction quality remains the same as long as the channel capacity is above the target rate. These two characteristics are known as the ``cliff effect''. Various joint source-channel coding schemes have been proposed in the literature to overcome the ``cliff effect'' \cite{Gunduz:IT:08,Kozintsev:TSP:98}, and to obtain graceful degradation of the signal quality with channel SNR, which typically combine multi-layer digital codes with multi-layer compression for unequal error protection.

In this paper we take a radically different approach,  and leverage the properties of uncoded transmission \cite{Goblick:IT:65, SoftCast:Allerton:10, Tung:CL:18} by directly mapping the real pixel values to  the complex-valued samples transmitted over the communication channel. Our goal is to design a JSCC scheme that bypasses the transformation of the pixel values to a sequence of bits, which are then mapped again to complex-valued channel inputs; and instead, directly maps the pixel values to channel inputs as in \cite{SoftCast:Allerton:10, Tung:CL:18}.

\section{DL-based JSCC}
\label{sec:jscc_algo}

Our design is inspired by the recent successful application of deep NNs (DNNs), and autoencoders, in particular, to the problem of source compression \cite{Balle:ICLR:17,TheisICLR2017, ChengPCS2018, AlexandreCVPR2018}, as well as by the first promising results in the design of end-to-end communication systems using autoencoder architectures \cite{OShea:ISSPIT:16, deep:PHY}.

The block diagram of the proposed JSCC scheme is shown in Fig. \ref{fig:comm_system2}. The encoder maps the $n$-dimensional input image $\bm x$ to a $k$-length vector of complex-valued channel input samples ${\bm z}$, which satisfies the average power constraint $\frac{1}{k} \mathbb{E}[{\bm z}^*{\bm z}] \leq P$, by means of a deterministic encoding function $f_{\bm{\theta}}: \mathbb{R}^n \rightarrow \mathbb{C}^k$. The encoder function $f_{\bm{\theta}}$ is parameterized using a CNN with parameters $\bm{\theta}$. The encoder CNN comprises a series of convolutional layers followed by parametric ReLU (PReLU) activation functions \cite{PReLU} and a normalization layer. The convolutional layers extract the image features, which are combined to form the channel input samples, while the nonlinear activation functions allow to learn a non-linear mapping from the source signal space to the coded signal space. The output $\tilde{\bm{z}} \in \mathbb{C}^k$ of the last convolutional layer of the encoder is normalized according to:
\begin{equation}
\bm z = \sqrt{kP}\frac{\tilde{\bm z}}{\sqrt{\tilde{\bm{z}}^*\tilde{\bm{z}}}}
\label{eq:power_constraint}
\end{equation}
where $\tilde{\bm{z}}^*$ is the conjugate transpose of $\tilde{\bm{z}}$, such that the channel input $\bm z$ satisfies the average transmit power constraint $P$.

Following the encoding operation, the joint source-channel coded sequence $\bm{z}$ is sent over the communication channel by directly transmitting the real and imaginary parts of the channel input samples over the I and Q components of the digital signal. The channel introduces random corruption to the transmitted symbols, denoted by $\eta : \mathbb{C}^k \rightarrow \mathbb{C}^k$. To be able to optimize the communication system in Fig. \ref{fig:comm_system2} in an end-to-end manner, the communication channel must be incorporated into the overall NN architecture. We model the communication channel as a series of non-trainable layers, which are represented by the transfer function $\hat{\bm z} = \eta({\bm z})$. We consider two widely used channel models: \textit{(i)} the AWGN channel, and \textit{(ii)} the slow fading channel. The transfer function of the Gaussian channel is $\eta_{n} (\bm z)= \bm z + \bm n$, where the vector $\bm n \in \mathbb{C}^k$ consists of independent identically distributed (i.i.d.) samples from a circularly symmetric complex Gaussian distribution, i.e., $ \bm n \sim  \mathcal{CN}(0,\sigma^2\bm I_k)$, where $\sigma^2$ is the average noise power. In the case of slow fading channel, we adopt the commonly used Rayleigh slow fading model. The multiplicative effect of the channel gain on the transmitted signal is captured by the channel transfer function $\eta_{h}(\bm z) =  h \bm z$, where $ h \sim \mathcal{CN}(0,H_c)$ is  a complex normal random variable. The joint effect of channel fading and Gaussian noise can be modelled by the composition of the transfer functions $\eta_h$ and $\eta_n$:  $\eta(\bm z) = \eta_n(\eta_h(\bm z))= h \bm z + \bm n$. Other channel models can be incorporated into the end-to-end system in a similar manner with the only requirement that the channel transfer function is differentiable in order to allow gradient computation and error back propagation.

The receiver comprises a joint source-channel decoder. The decoder maps the corrupted complex-valued signal $\hat{\bm{z}} = \eta(\bm{z}) \in \mathbb{C}^k$ to an estimation of the original input $ \hat{\bm{x}} \in \mathbb{R}^n$ using a decoding function $g_{\bm{\phi}}: \mathbb{C}^k \rightarrow \mathbb{R}^n$. Similarly to the encoding function, the decoding function is parameterized by the decoder CNN with parameter set ${\bm{\phi}}$. The NN decoder inverts the operations performed by the encoder by passing the received (and possibly corrupted) coded signal $\hat{\bm z}$ through a series of transpose convolutional layers (with non linear activation functions) in order to map the image features to an estimate $\hat{\bm x}$ of the originally transmitted image.

The encoding and decoding functions are designed jointly to minimize the average distortion between the original input image $\bm x$ and its reconstruction $\hat{\bm x}$ produced by the decoder:
\begin{equation}
(\bm{\theta}^*, \bm{\phi}^*)=\argmin_{{\bm \theta}, \bm{\phi}} \mathbb{E}_{p(\bm x, \hat{\bm x})}[d(\bm x, \hat{\bm x})],
\label{eq:exp_distortion}
\end{equation}
where $d(\bm x, \hat{\bm x})$ is a given distortion measure, and $p(\bm x, \hat{\bm x})$ is the joint probability distribution of the original and reconstructed images. Since the true distribution of the input data $p(\bm x)$ is often unknown, an analytical form of the expected distortion in Eq. \eqref{eq:exp_distortion} is also unknown. We, therefore, estimate the expected distortion by sampling from an available dataset.

\begin{figure}[t]
	\begin{center}
		\includegraphics[width = 0.5\textwidth]{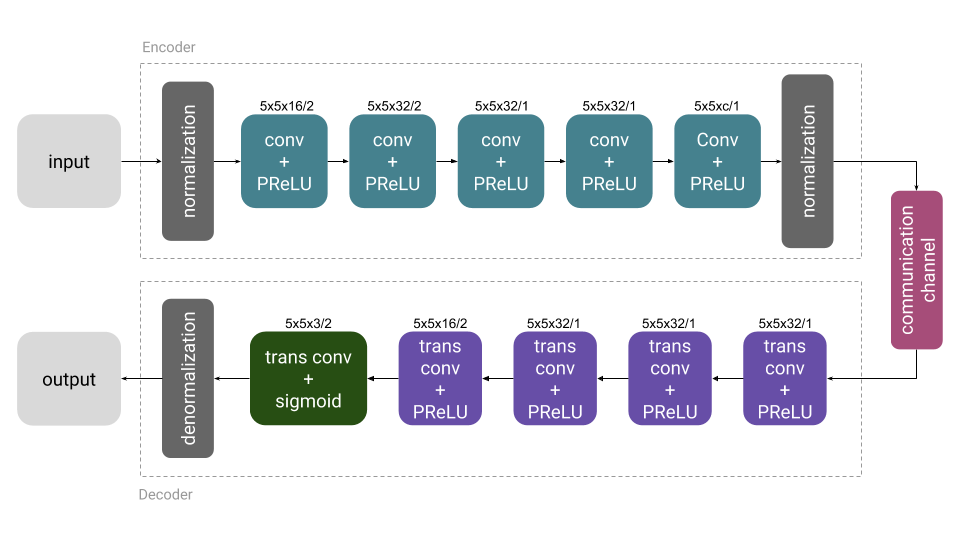}
	\end{center}
		\caption{Encoder and decoder NN architectures used in the implementation of the proposed deep JSCC scheme.}\label{fig:architecture}
\end{figure}

\section{Evaluation}\label{s:evaluation}

To demonstrate the potential of our proposed deep JSCC scheme, we use the NN architecture depicted in Fig. \ref{fig:architecture}. At the encoder, the normalization layer is followed by five convolutional layers. Since the statistics of the input data are generally not known at the decoder, the input images are normalized by the maximum pixel value $255$, producing pixel values in the $[0,1]$ range. The notation $F\times F \times K/S$ denotes a convolutional layer with $K$ filters of spatial extent (or size) $F$ and stride $S$. The values of the hyperparameters $F,K$ and $S$ used in our experiments are given in Fig. \ref{fig:architecture}. PReLU activation function is applied to the output of all convolutional layers. The output of the last convolutional layer, which consists of $2k$ units, is followed by another normalization layer which enforces the average power constraint specified in Eq. \eqref{eq:power_constraint}. The output of the normalization layer is combined into $k$ complex-valued channel input samples and forms the encoded signal representation, which is transmitted over the channel.

The decoder inverts the operations performed by the encoder. The real and imaginary parts of the $k$ complex-valued noisy channel output samples are combined into $2k$ values which are fed into the transpose convolutional layers. The latter progressively transform the corrupted image features into an estimation of the original input image, while upsampling it to the correct resolution. The hyperparameters of the decoder layers mirror the corresponding values of the encoder layers (Fig. \ref{fig:architecture}). The output of all transpose convolutional layers of the decoder except for the last one are passed through a PReLU activation function, while a sigmoid nonlinearity is applied to the output of the last transpose convolutional layer in order to produce values in the $[0,1]$ range. Finally, a denormalization layer multiplies the output values by $255$ in order to generate pixel values within the $[0,255]$ range.

The above architecture is implemented in Tensorflow \cite{tensorflow2015-whitepaper}. We use the Adam optimization framework \cite{KingmaARXIV2014}, which is a form of stochastic gradient descent. Our loss function is the average mean squared error (MSE) between the original input image $\bm x$  and the reconstruction $\hat{\bm x}$  at the output of the decoder, defined as:
\begin{equation}
\mathcal{L} = \frac{1}{N}\sum_{i=1}^N d(\bm x_i, \hat{\bm x_i}),
\end{equation}
where $d(\bm x, \hat{\bm x})=\frac{1}{n}||\bm x-\hat{\bm x} ||^2$ is the mean squared-error distortion and $N$ is the number of samples. In order to achieve various bandwidth compression ratios $k/n$, we vary the number of filters $K$ in the last convolutional layer of the encoder. Since our architecture is fully convolutional, it can be trained and deployed on input images of any resolution.

The performance of the deep JSCC algorithm, as well as of all benchmark schemes is quantified in terms of $\mathrm{PSNR}$. The PSNR metric measures the ratio between the maximum possible power of the signal and the power of the noise that corrupts the signal. The PSNR is defined as follows:
\begin{equation}
\mathrm{PSNR}=10\log_{10}\frac{\mathrm{MAX}^2}{\mathrm{MSE}}~ \mathrm{~(dB)}.
\label{eq:psnr}
\end{equation}
where $\mathrm{MSE}=d(\bm{x},\hat{\bm{x}})$ is the mean squared-error between the reference image $\bm{x}$ and the reconstructed image $\hat{\bm{x}}$, and $\mathrm{MAX}$ is the maximum possible value of the image pixels. All our experiments are conducted on 24-bit depth RGB images (8 bits per pixel per colour channel), thus $\mathrm{MAX} = 2^8-1=255$.

The channel SNR is defined as:
\begin{equation}
\mathrm{SNR}=10\log_{10}\frac{P}{\sigma^2}~ \mathrm{~(dB)},
\label{eq:snr}
\end{equation}
and represents the ratio of the average power of the coded signal (channel input signal) to the average noise power. Recall that $P$ is the average power of the channel input signal after applying the power normalization layer at the encoder of the proposed JSCC scheme. For benchmark schemes that use explicit signal modulation, $P$ is the average power of the symbols in the constellation. Without loss of generality, we set the average signal power  to $P=1$ for all experiments.

\subsection{Evaluation on CIFAR-10 dataset}
\label{sec:cifar10_eval}

We start by evaluating our deep JSCC scheme on the CIFAR-10 image dataset. The training data consists of  $50000$ $32\times32$ training images \cite{CIFARdataset} combined with random realizations of the channel under consideration. The performance of the proposed JSCC scheme is tested on $10000$ test images from the CIFAR-10 dataset, which are distinct from the images used for training. We initially set the learning rate to $10^{-3}$ and reduce it after 500k iterations to $10^{-4}$. We use a mini-batch size of $64$ samples and train our models until the performance on the test set does not improve further. However, we would like to emphasize that we do not use the test set images to optimize the network hyperparameteres. During performance evaluation we transmit each image 10 times in order to mitigate the effect of randomness introduced by the communication channel.

We first investigate the performance of our proposed deep JSCC algorithm in the AWGN setting, i.e., the channel transfer function is $\eta=\eta_{n}$. We vary the SNR by varying the noise variance $\sigma^2$ and compare the proposed deep JSCC algorithm with an upper bound on any digital transmission scheme, which employs JPEG or JPEG2000 for source compression. The computation of the upper bound is based on the Shannon's separation theorem, which states that the necessary and sufficient condition for reliable communication over a discrete memoryless channel with channel capacity $C$ is
\begin{equation}
nR \leq kC.
\label{eq:shannon_bound}
\end{equation}
The above expression defines the maximum rate
\begin{equation}
R_{\mathrm{max}} = \frac{k}{n}C
\label{eq:max_rate}
\end{equation}
for a channel with capacity $C$ at which the source can be compressed and transmitted with arbitrarily small probability of error.  Thus, to compute the upper bound, we first compute the maximum number of bits per source sample $R_\mathrm{max}$ using Eq. \eqref{eq:max_rate}, where $C=\log_2(1+\mbox{SNR})$ for a complex AWGN channel. This is the maximum rate for source compression that is guaranteed reliable transmission over the channel. Since JPEG and JPEG2000 cannot compress the image data at an arbitrarily low bitrate, we also compute the minimum bitrate value $R_\mathrm{min}$ beyond which compression results in complete loss of information and the original image cannot be reconstructed. If, for a given set of values of $n$, $k$ and $C$, the minimum rate $R_\mathrm{min}$ exceeds the maximum allowable rate $R_\mathrm{max}$,  we assume that the image cannot be reliably transmitted and each color channel is reconstructed to the mean value of all the pixels for that channel. When $R_\mathrm{min} < R_\mathrm{max}$, we compress the images at the largest rate $R$ that satisfies $R \leq R_\mathrm{max}$ (since, again, it is not always possible to achieve an arbitrary target bitrate $R_\mathrm{max}$ with JPEG or JPEG2000 compression software), and measure the distortion between the reference image and the compressed one, assuming that the compressed bitstream can be transmitted without errors.

We would like to note that we do not use any explicit practical channel coding and modulation scheme in the computation of the bound. Compressing the source at rate $R_\mathrm{max}$ and assuming error-free transmission at this rate, implicitly suggests that one would need to use a capacity-achieving combination of channel code and modulation scheme to achieve reliable transmission. Thus, the performance of any digital transmission scheme that employs an actual channel coding scheme and modulation along with JPEG/JPEG2000 compression will be inferior to this upper bound.

\begin{figure}[t]
	\begin{center}
 		{\includegraphics[width=0.5 \textwidth]{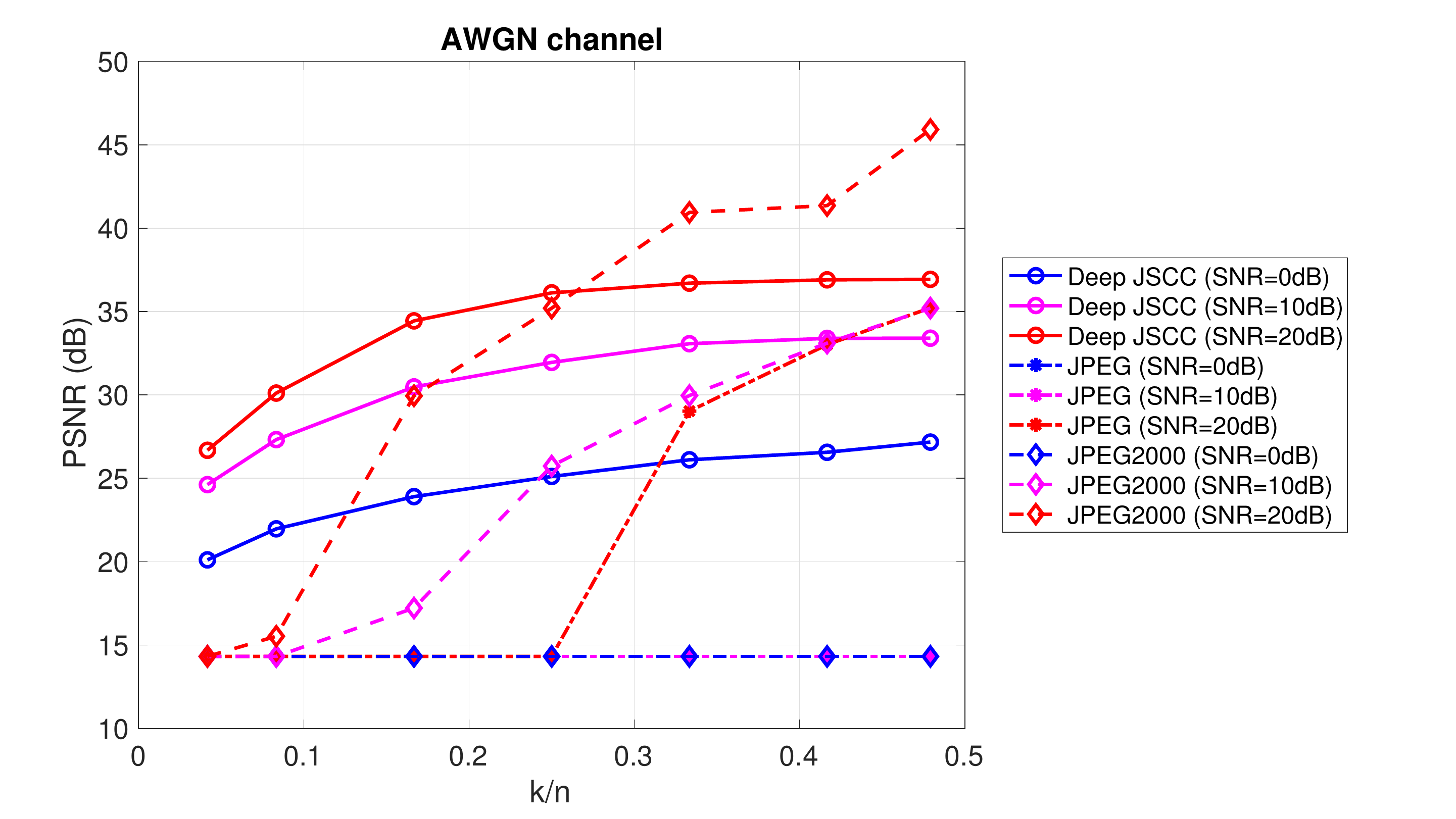}}
	\end{center}
		\caption{Performance of the deep JSCC algorithm on CIFAR-10 test images over an AWGN channel with respect to the compression ratio, $k/n$, for different $\mathrm{SNR}$ values. For each case, the same $\mathrm{SNR}$ value is used in training and evaluation.} \label{fig:cifar10_awgn_PSNR_vs_CR}
\end{figure}

Fig. \ref{fig:cifar10_awgn_PSNR_vs_CR}  illustrates the performance of the proposed deep JSCC algorithm with respect to the bandwidth compression ratio, $k/n$, in different SNR regimes. This performance is compared against the upper bound on the performance of any digital scheme that employs JPEG/JPEG2000 for compression.  We note that the threshold behavior of the upper bound in the figure is not due to the ``cliff effect''. The initial flat part of these curves is due to the fact that JPEG and JPEG2000 completely break down in this region, i.e., the maximum transmission rate $R_{max}$ is below the minimum number of bits per pixel, $R_{min}$, required to compress the images at the worst quality and obtain a meaningful reconstruction at the decoder.

We observe that, in very bad channel conditions (e.g., for SNR=0dB), the digital schemes deploying JPEG or JPEG2000 would break down, while with the proposed deep JSCC scheme transmission is possible with reasonably good performance.  At medium and high SNRs and for limited channel bandwidth, i.e., for $k/n \in [0.04,0.2]$, the performance of the proposed deep JSCC scheme is considerably above the one that can be achieved by JPEG and JPEG2000 even assuming that reliable transmission at channel capacity is possible\footnote{While near capacity-achieving channel codes exist for the AWGN channel, these typically require very large blocklengths. It is known that the achievable rates guaranteeing a low block error probability for the blocklengths considered here are below the capacity \cite{Polyanskiy:IT:10} for the entire range of compression ratio values. Therefore, the upper bounds in Fig. \ref{fig:cifar10_awgn_PSNR_vs_CR} are typically not achievable.}. Even when the channel bandwidth becomes less constrained, i.e., for $k/n> 0.3$, the performance of the deep JSCC scheme remains competitive with its JPEG/JPEG2000 counterparts. The saturation of the proposed deep JSCC scheme in the large channel bandwidth regime is possibly due to the limited capability of the particular autoencoder architecture employed, which may be improved, for example, by employing a different activation function than PReLU as in \cite{Balle:ICLR:17}, or through incremental training as in \cite{TheisICLR2017}.

\begin{figure}[t!]
	\begin{center}
 		\subfloat[]{\includegraphics[width=0.5\textwidth]{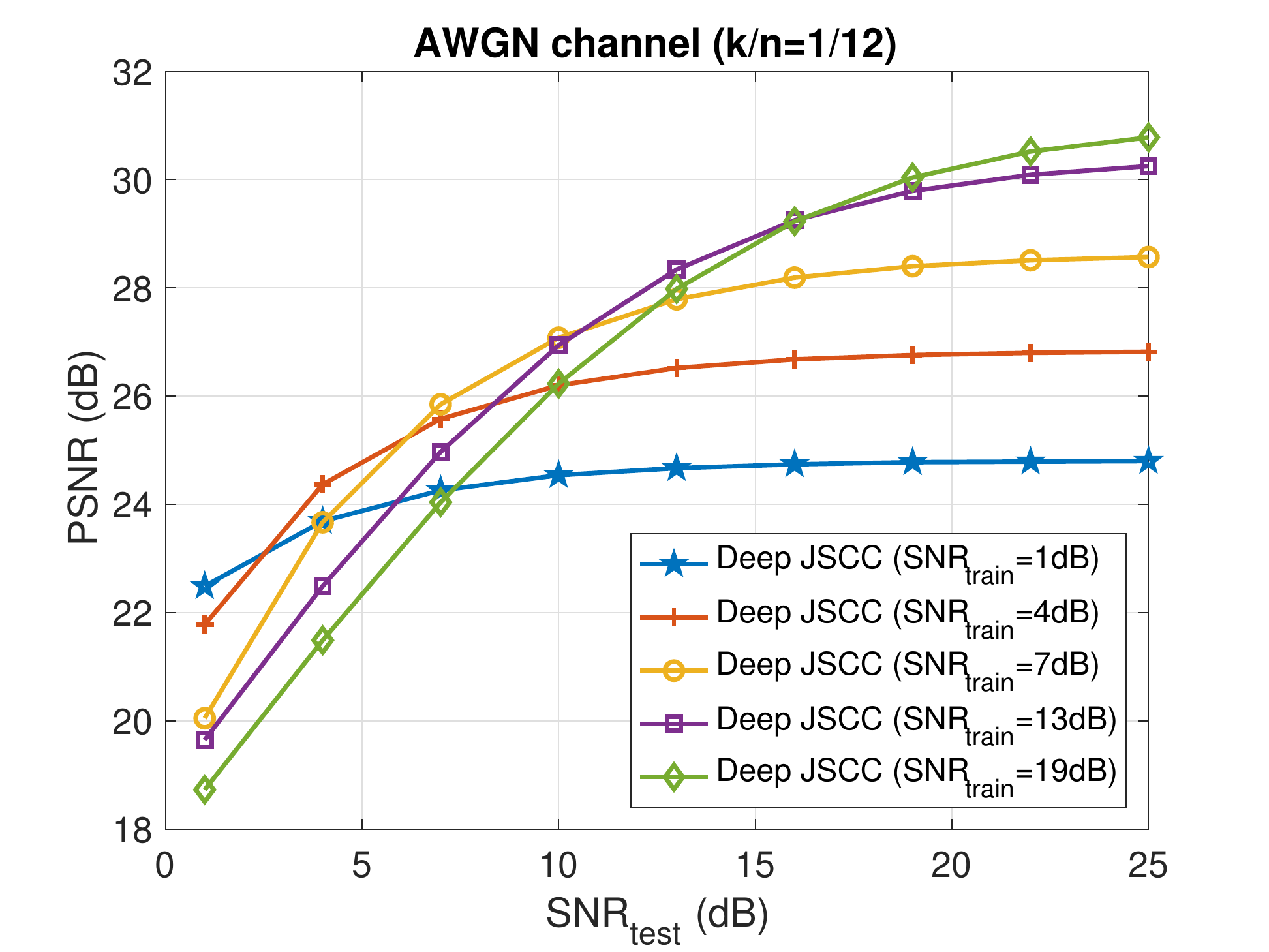}\label{fig:cifar10_awgn_1over12}} \\
		\subfloat[]{\includegraphics[width= 0.5\textwidth]{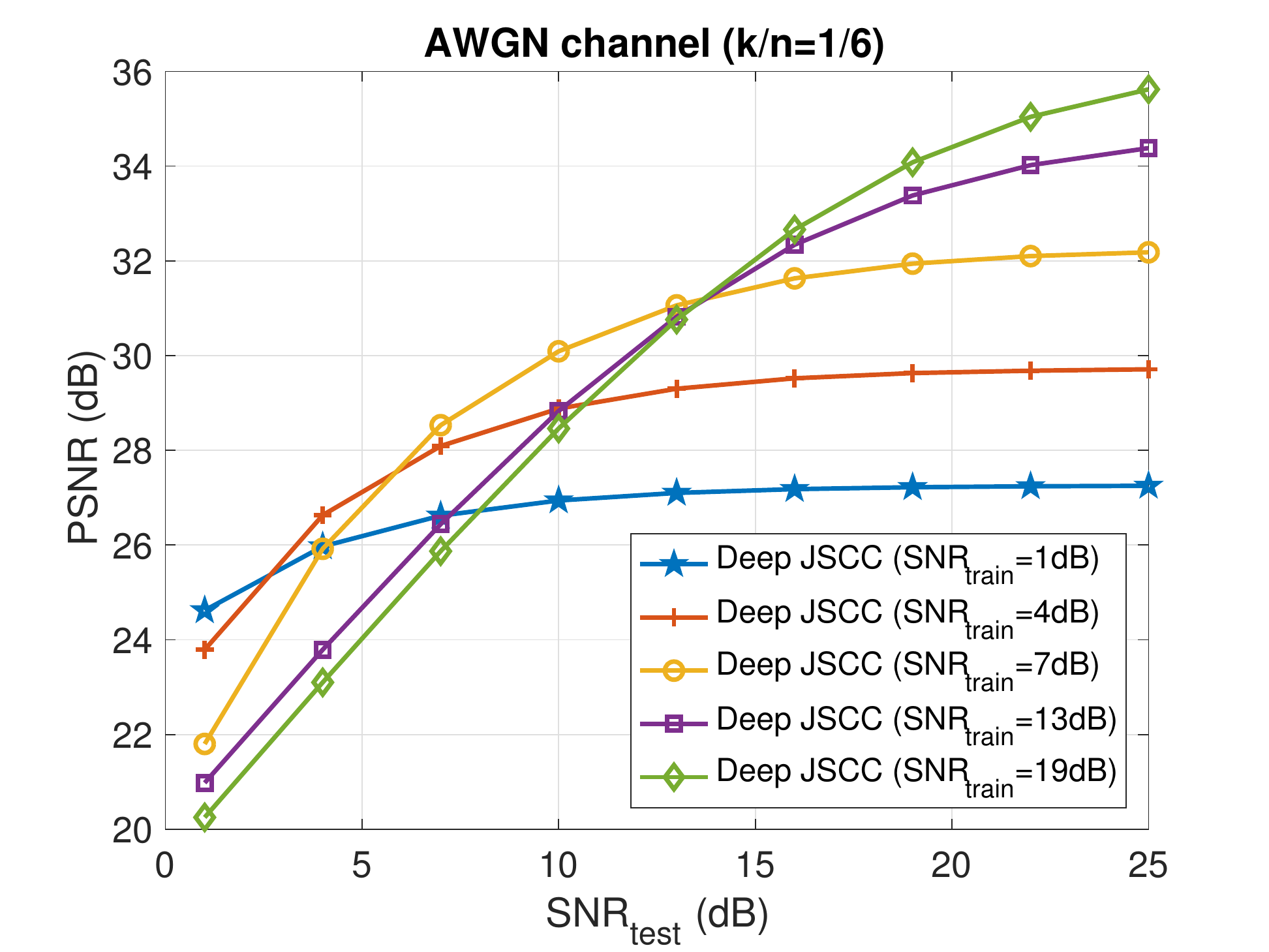}\label{fig:cifar10_awgn_1over6}}
	\end{center}
		\caption{Performance of the deep JSCC algorithm on CIFAR-10 test images with respect to the channel SNR over an AWGN channel for bandwidth compression ratios (a) $k/n=1/12$ and (b) $k/n=1/6$. Each curve is obtained by training the encoder/decoder network for a particular channel SNR value.}
	\label{fig:cifar10_awgn_psnr_vs_snr}
\end{figure}

We next study the robustness of the proposed deep JSCC scheme to variations in channel conditions. Figs. \ref{fig:cifar10_awgn_1over12} and \ref{fig:cifar10_awgn_1over6} illustrate the average $\mathrm{PSNR}$ of the reconstructed images versus the $\mathrm{SNR}$ of the AWGN channel for two different values of bandwidth compression ratio, $k/n$. Each curve in Figs. \ref{fig:cifar10_awgn_1over12} and \ref{fig:cifar10_awgn_1over6} is generated by training our end-to-end system for a specific channel $\mathrm{SNR}$ value, denoted as $\mathrm{{SNR}_{train}}$, and then evaluating the performance of the learned encoder/decoder parameters on the test images for varying $\mathrm{SNR}$ values, denoted as $\mathrm{{SNR}_{test}}$. In other words, each curve represents the performance of the proposed JSCC scheme optimized for channel SNR  equal to $\mathrm{{SNR}_{train}}$, and deployed in different channel conditions with SNR equal to $\mathrm{{SNR}_{test}}$. These results provide an insight into the performance of the proposed algorithm when the channel conditions are different from those for which the end-to-end system is optimized and demonstrate the robustness of the proposed JSCC to variations in channel quality. We can observe that for $\mathrm{{SNR}_{test}} < \mathrm{{SNR}_{train}}$, i.e., when the channel conditions are worse than those for which the encoder/decoder have been optimized, our deep JSCC algorithm does not suffer from the ``cliff effect'' observed in digital systems. Unlike digital systems, where the quality of the decoded signal drops sharply when $\mathrm{{SNR}_{test}}$ drops below a critical threshold value, the deep JSCC scheme is more robust to channel quality fluctuations and exhibits a gradual performance degradation as the channel deteriorates. Such behavior is akin to the performance of an analog scheme \cite{Goblick:IT:65, Gunduz:IT:08, Tung:CL:18}, and is attributed to the capability of the autoencoder to map similar images/features to nearby points in the channel input signal space; thus, with decreasing $\mathrm{{SNR}_{test}}$ the decoder can still obtain a reconstruction of the original image.

On the other hand, when $\mathrm{{SNR}_{test}}$ increases above $\mathrm{{SNR}_{train}}$, we observe initially a gradual improvement in the quality of the reconstructed images before the performance finally saturates as $\mathrm{{SNR}_{test}}$ increases beyond a certain value. The performance in the saturation region is driven solely by the amount of compression implicitly decided during the training phase for the target value $\mathrm{{SNR}_{train}}$. It is worth noting that performance saturation does not occur at $\mathrm{{SNR}_{test}} = \mathrm{{SNR}_{train}}$ as in digital image/video transmission systems \cite{SoftCast:Allerton:10}, but at $\mathrm{{SNR}_{test}} > \mathrm{{SNR}_{train}}$. This behavior indicates that the proposed JSCC scheme determines an implicit trade-off between the amount of error protection and compression, which does not necessarily target an error-free transmission when the system operates at $\mathrm{{SNR}_{test}}= \mathrm{{SNR}_{train}}$. We also note that when the encoder/decoder are optimized for very high $\mathrm{{SNR}_{train}}$, and $\mathrm{{SNR}_{test}}> \mathrm{{SNR}_{train}}$, the system boils down to an ordinary autoencoder, and its performance is solely limited by the degree-of-freedom imposed by the bandwidth compression ratio $k/n$, i.e., the dimension of the bottleneck layer of the autoencoder.

Next we study the performance of our deep JSCC scheme under the assumption of a slow Rayleigh fading channel with AWGN. In this case,  the channel transfer function is $\eta(\bm z)=h\bm z + \bm n$, where $ h \sim \mathcal{CN}(0,H_c)$ and $ \bm n \sim  \mathcal{CN}(0,\sigma^2\bm I_k)$. In this experiment, we do not assume channel state information either at the receiver or the transmitter, or consider the transmission of pilot signals. As we assume slow fading, the channel gain $h$ is randomly sampled from the complex Gaussian distribution $ \mathcal{CN}(0,H_c)$ for each transmitted image and remains constant during the transmission of the entire image, and changes independently to another state for the next image. We set $H_c =1$ and vary the noise variance $\sigma^2$ to emulate varying average channel SNR.

\begin{figure}[t]
	\begin{center}
 		\includegraphics[width= 0.5\textwidth]{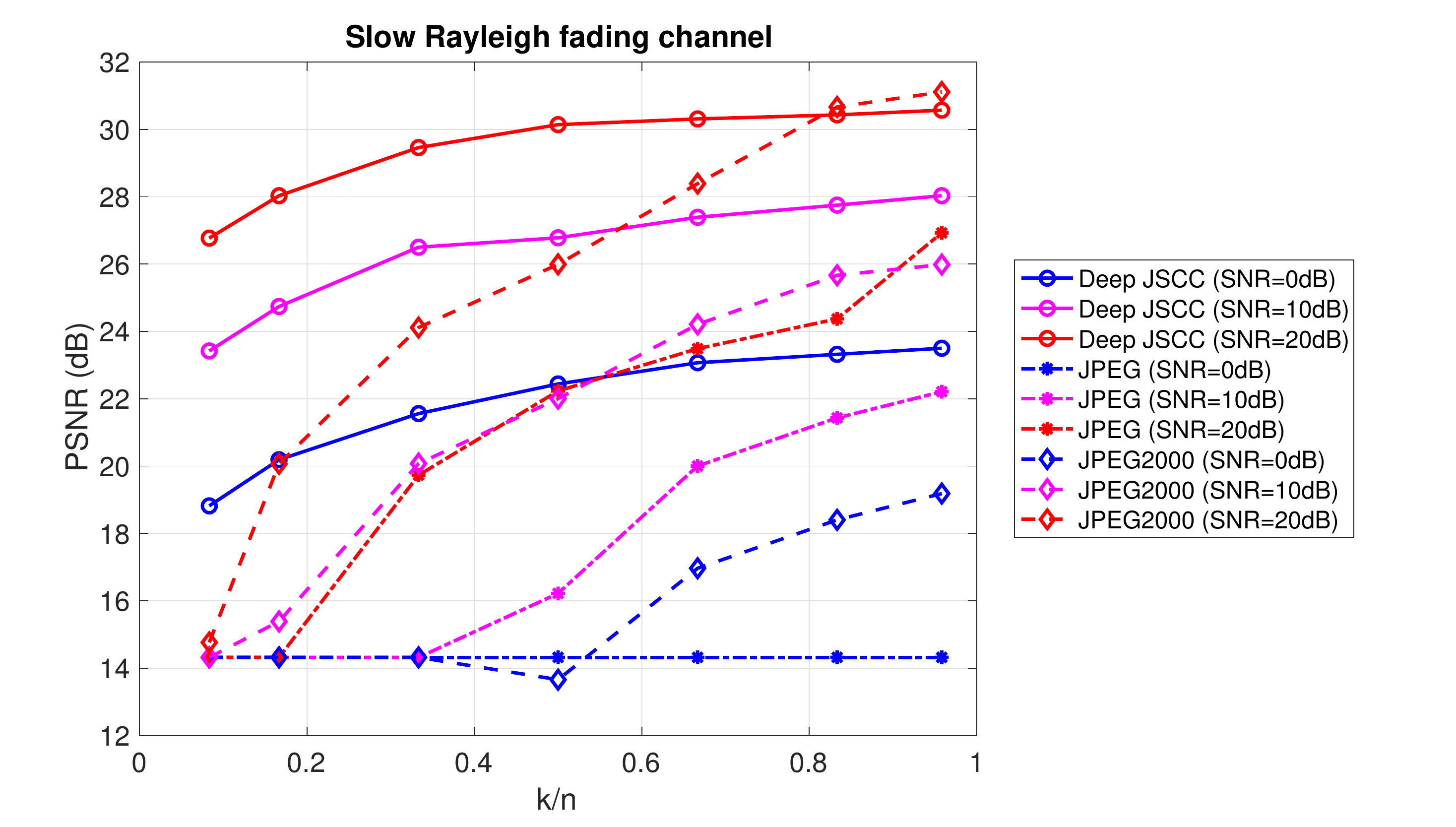}
	\end{center}
		\caption{Performance of the deep JSCC algorithm on CIFAR-10 test images over a slow Rayleigh fading channel with respect to the bandwidth compression ratio, $k/n$, for different $\mathrm{SNR}$ values. For each case, the same target $\mathrm{SNR}$ value is used in training and evaluation.}
		\label{fig:cifar10_fading_PSNR_vs_CR}
\end{figure}

In Fig. \ref{fig:cifar10_fading_PSNR_vs_CR}, we plot the performance of the proposed deep JSCC algorithm over a slow Rayleigh fading channel as a function of the bandwidth compression ratio, $k/n$, for different average $\mathrm{SNR}$ values. Note that, due to the lack of channel state information, the capacity of this channel in the Shannon sense is zero, since no positive rate can be guaranteed reliable transmission at all channel conditions; that is, for any positive transmission rate, the channel capacity will be below the transmission rate with a non-zero probability. Therefore, we calculate an upper bound on any digital transmission scheme designed for the average SNR value. i.e., for $\mathrm{SNR}=10\log_{10}\frac{\mathbb{E}[h^2]P}{\sigma^2}$, which uses JPEG/JPEG2000 for compression. Similarly to the case of the AWGN channel, we assume that the source image is compressed with JPEG/JPEG2000 at rate that is equal to the capacity of the complex AWGN channel at the average SNR value. That is, we calculate the maximum number of bits that can be transmitted reliably using Eq. \eqref{eq:max_rate}, where the channel capacity is calculated for the average SNR value.  If the channel capacity is below this value due to fading, an outage occurs, and the mean pixel values are used for reconstruction, i.e., maximum distortion is reached. If the channel capacity is above the transmission rate, the transmitted codeword can be decoded reliably.  We observe that deep JSCC beats the upper bound on the digital transmission schemes at all SNR and bandwidth compression values. This result emphasizes the benefits of the proposed deep JSCC technique when communicating over a  time-varying channel, or multicasting to multiple receivers with varying channel states.

\begin{figure}[t!]
	\begin{center}
		\subfloat[]{\includegraphics[width= 0.5\textwidth]{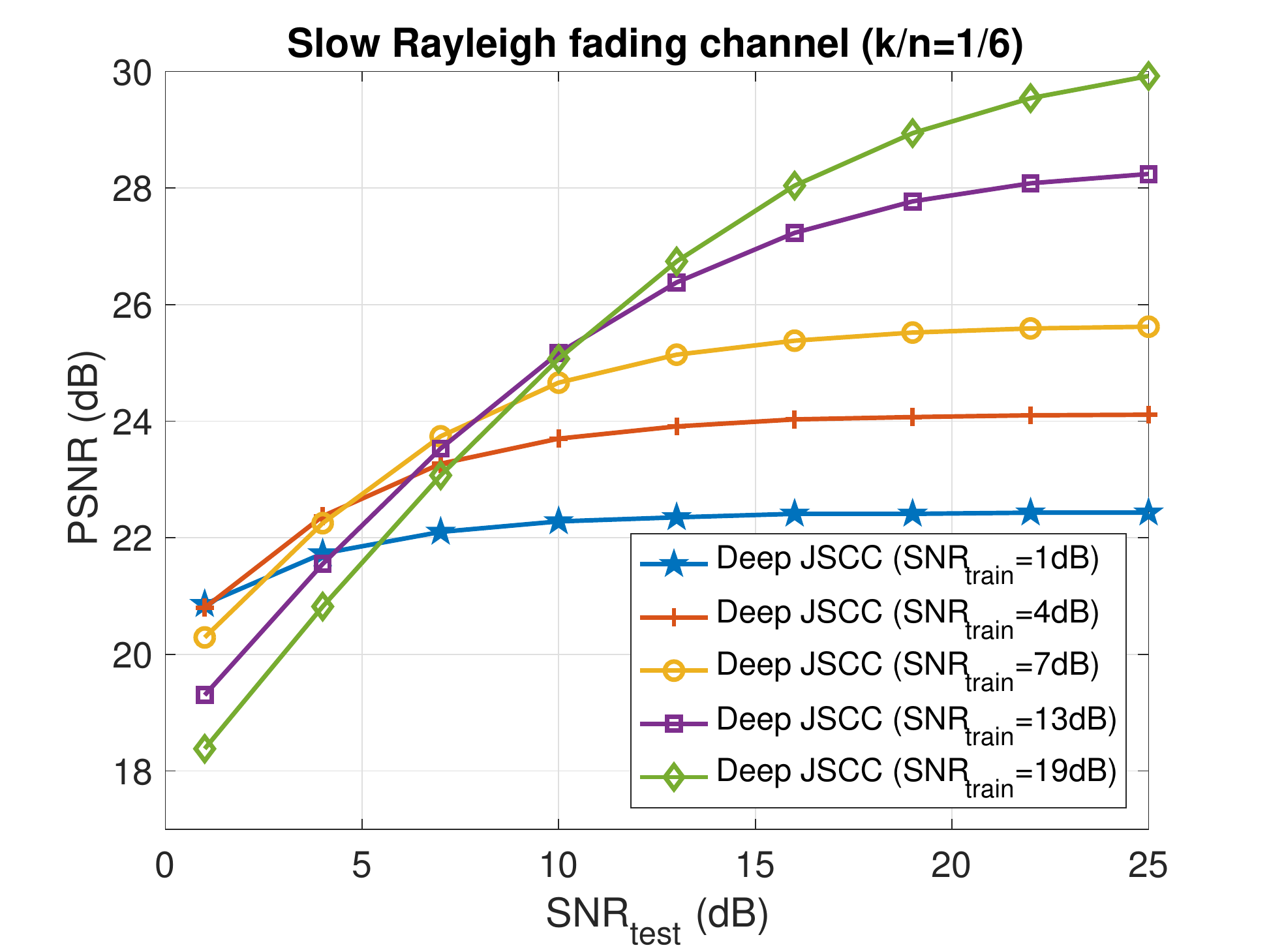}\label{fig:cifar10_fading_1over6}} \\
		\subfloat[]{\includegraphics[width=0.5\textwidth]{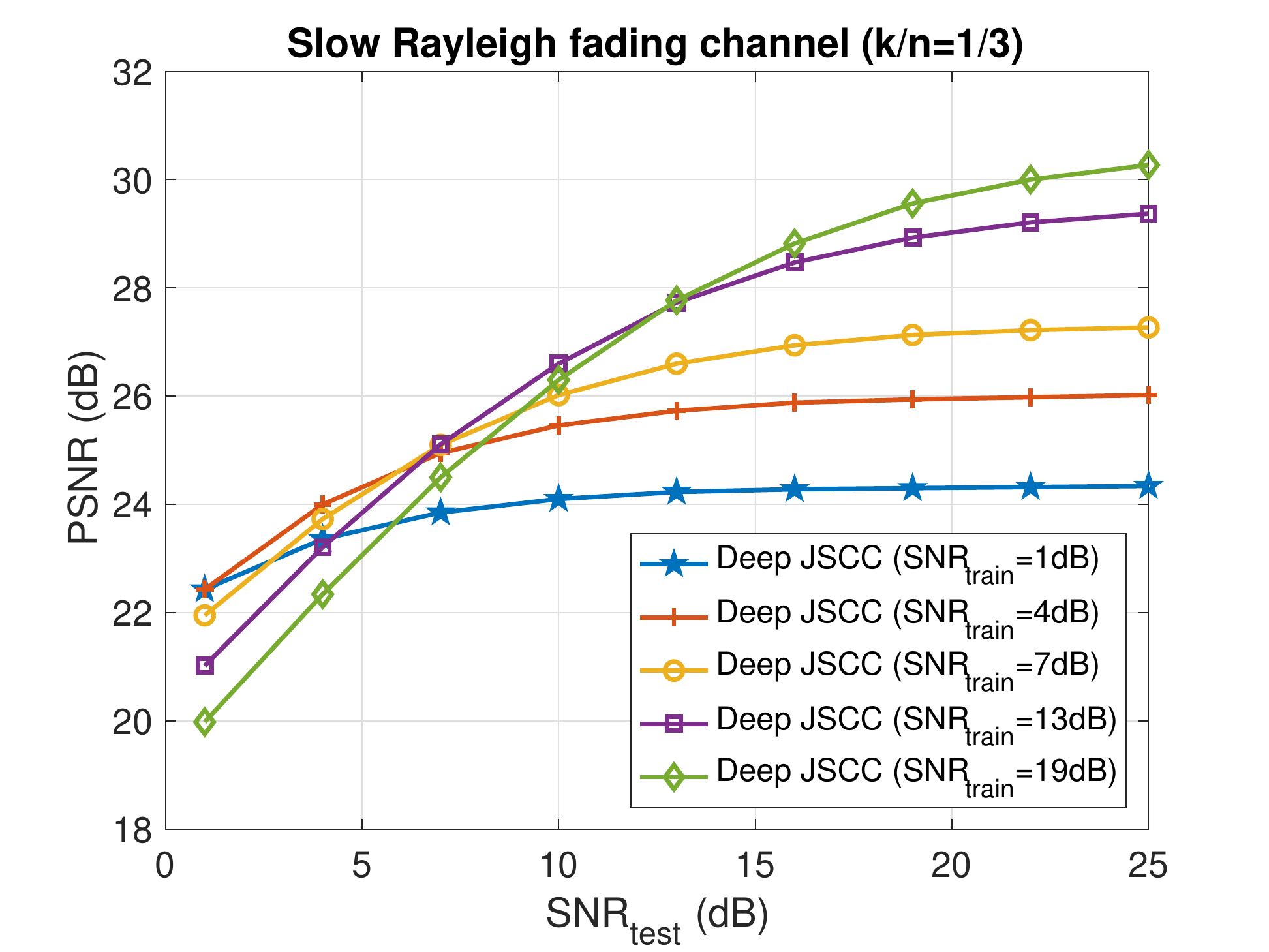}\label{fig:cifar10_fading_1over3}}
	\end{center}
		\caption{Performance of the deep JSCC algorithm on CIFAR-10 test images with respect to the average channel SNR over an AWGN slow Rayleigh fading channel for bandwidth compression ratios (a) $k/n=1/6$ and (b) $k/n=1/3$. Each curve is obtained by training the encoder/decoder network for a particular channel SNR value.}
	\label{fig:fading_psnr_vs_snr}
\end{figure}

We  illustrate the robustness of the proposed deep JSCC scheme to variations of the average channel SNR in a slow Rayleigh fading channel in Figs. \ref{fig:cifar10_fading_1over6} and \ref{fig:cifar10_fading_1over3}. We observe that, while the performance of the deep JSCC scheme drops compared to the static AWGN channel, the quality of the reconstructed images is still reasonable, despite the lack of channel state information. This suggests that the network learns to estimate the channel state, and adapts the decoder accordingly; that is, the proposed deep JSCC scheme combines not only source coding, channel coding, and modulation, but also channel estimation, into one single component, whose parameters are learned through training.

\subsection{Evaluation on the Kodak dataset}
\label{sec:kodak}

We also evaluate the proposed deep JSCC scheme on higher resolution images. To this end, we train our NN architecture on the Imagenet  dataset \cite{imagenet_cvpr09} which consists of $1.2$ million images. The images are randomly cropped to patches of size $128 \times 128$ and fed into the network in mini-batches of $32$ samples. We set the learning rate to $10^{-4}$ and train the models until convergence. The evaluation is performed on the Kodak image dataset\footnote{http://r0k.us/graphics/kodak/} consisting of 24 $768\times512$ images. During evaluation, each image is transmitted $100$ times, so that the performance can be averaged over multiple realizations of the random channel.

We first investigate the performance of the proposed deep JSCC algorithm over an AWGN channel by varying the noise power $\sigma^2$. The performance of the proposed deep JSCC algorithm is compared against digital transmission schemes that use JPEG/JPEG2000 for image compression followed by practical channel coding and modulation schemes. We use all possible combinations of $(4096,8192)$, $(4096,6144)$, and $(2048,6144)$ LDPC codes (which correspond to $1/2$, $2/3$ and $1/3$ rate codes) with BPSK, 4-QAM, 16-QAM and 64-QAM digital modulation schemes. For the sake of legibility, we only present the best performing digital transmission schemes and omit those that perform similarly, or whose performance in terms of PSNR is below 15dB.

\begin{figure}[t]
	\begin{center}
		\subfloat[ ]{\includegraphics[width= 0.5\textwidth]{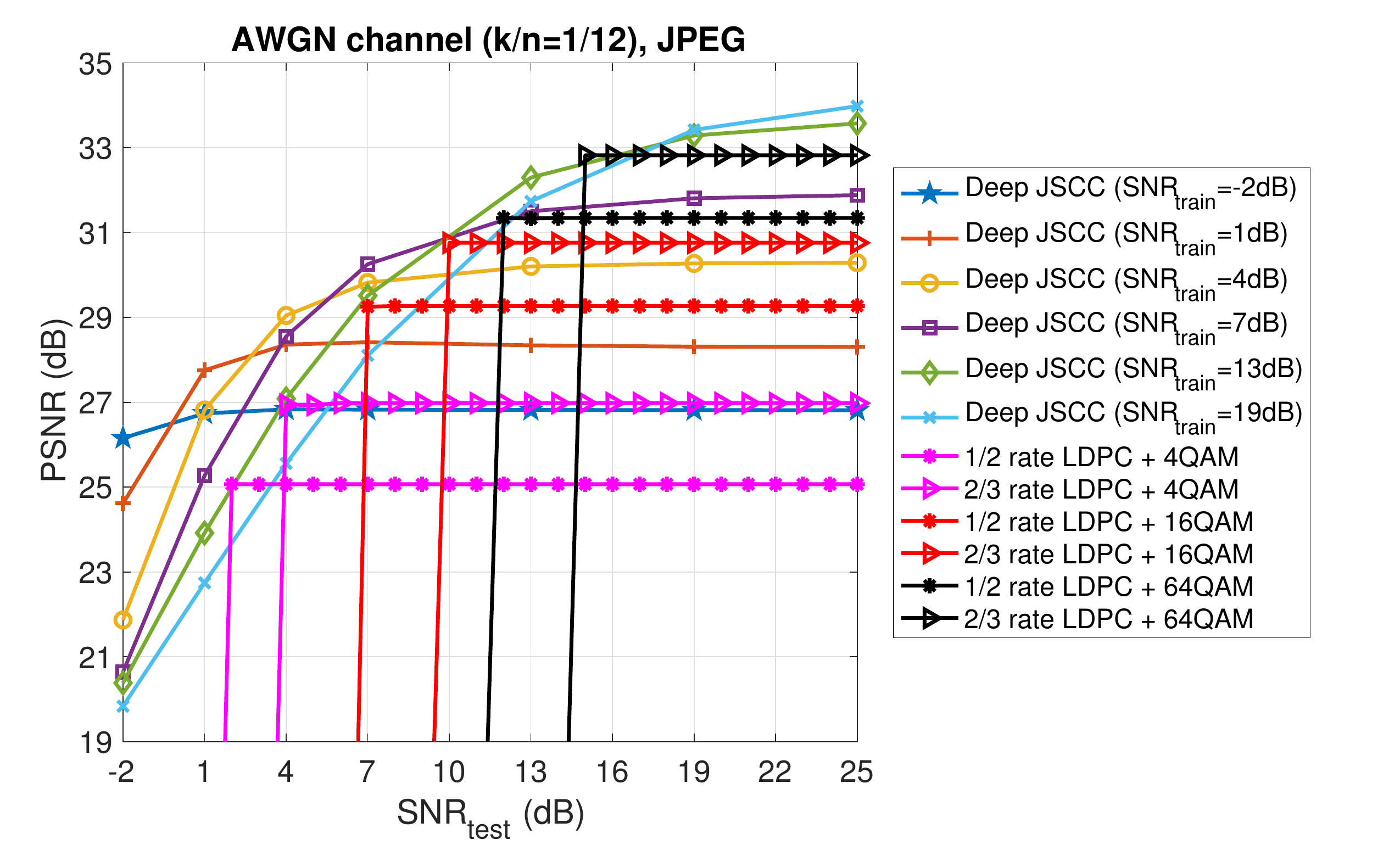}\label{fig:kodak_awgn_1over12_jpeg}} \\
		\subfloat[ ]{\includegraphics[width=0.5\textwidth]{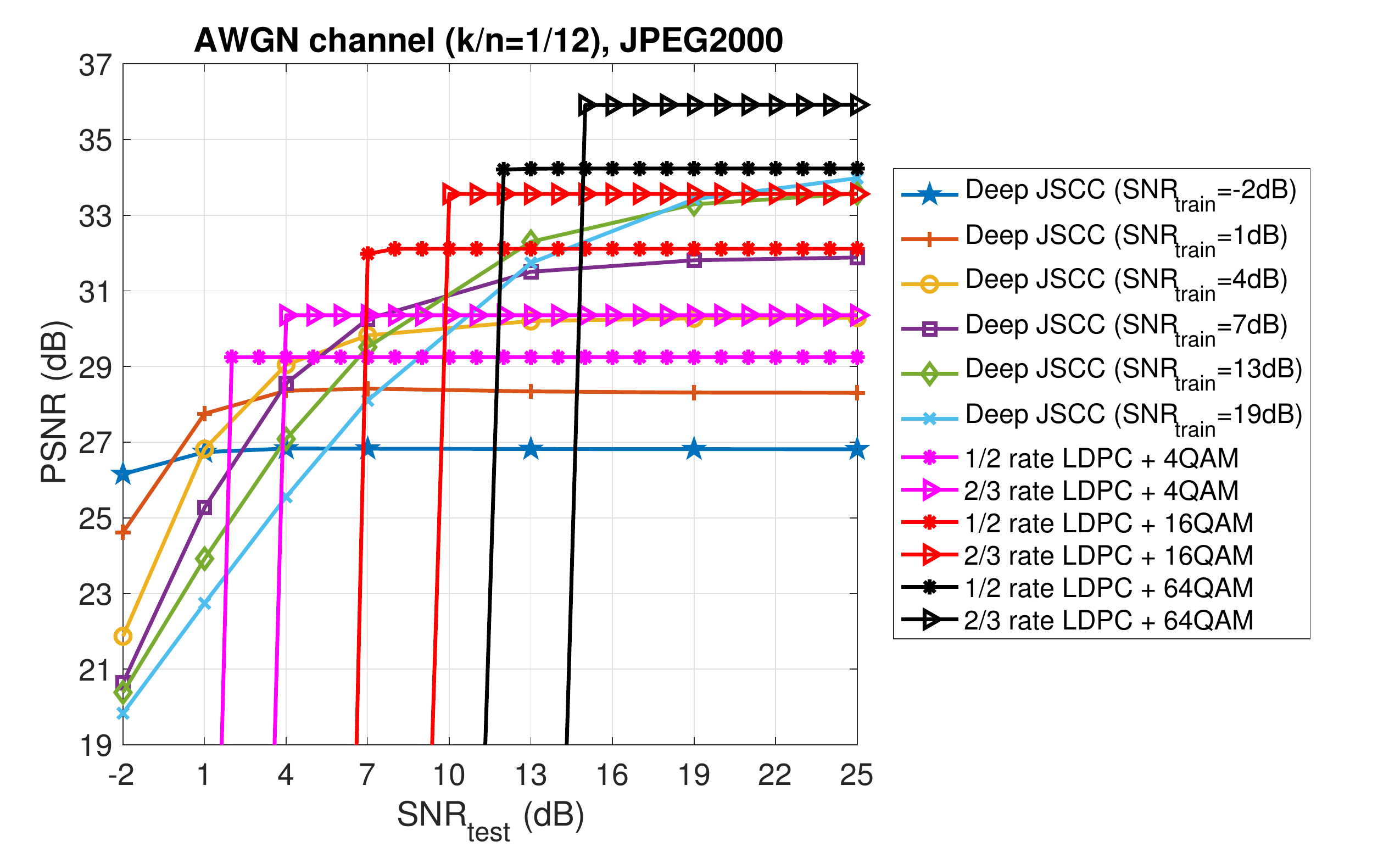}\label{fig:kodak_awgn_1over12_jpeg2000}}
	\end{center}
		\caption{Performance comparison of deep JSCC with baseline digital transmission schemes on the Kodak image dataset over AWGN channel for bandwidth compression ratio $k/n=1/12$. The digital schemes employ (a) JPEG and (b) JPEG2000 for image compression and various channel codes and modulation schemes. }
	\label{fig:kodak_awgn_1over12}
\end{figure}

\begin{figure}[t]
	\begin{center}
		\subfloat[]{\includegraphics[width= 0.5\textwidth]{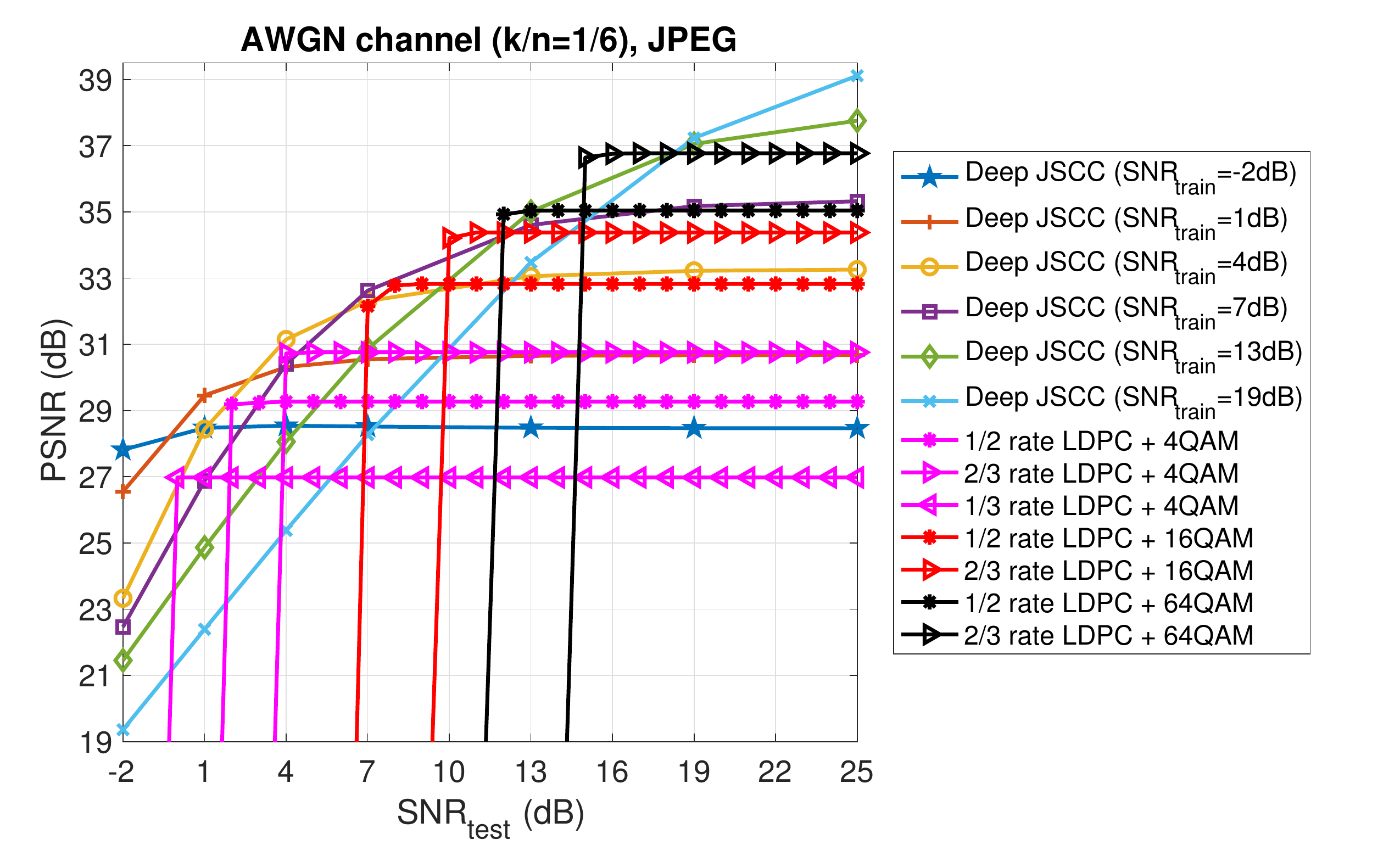}\label{fig:kodak_awgn_1over6_jpeg}} \\
		\subfloat[]{\includegraphics[width=0.5\textwidth]{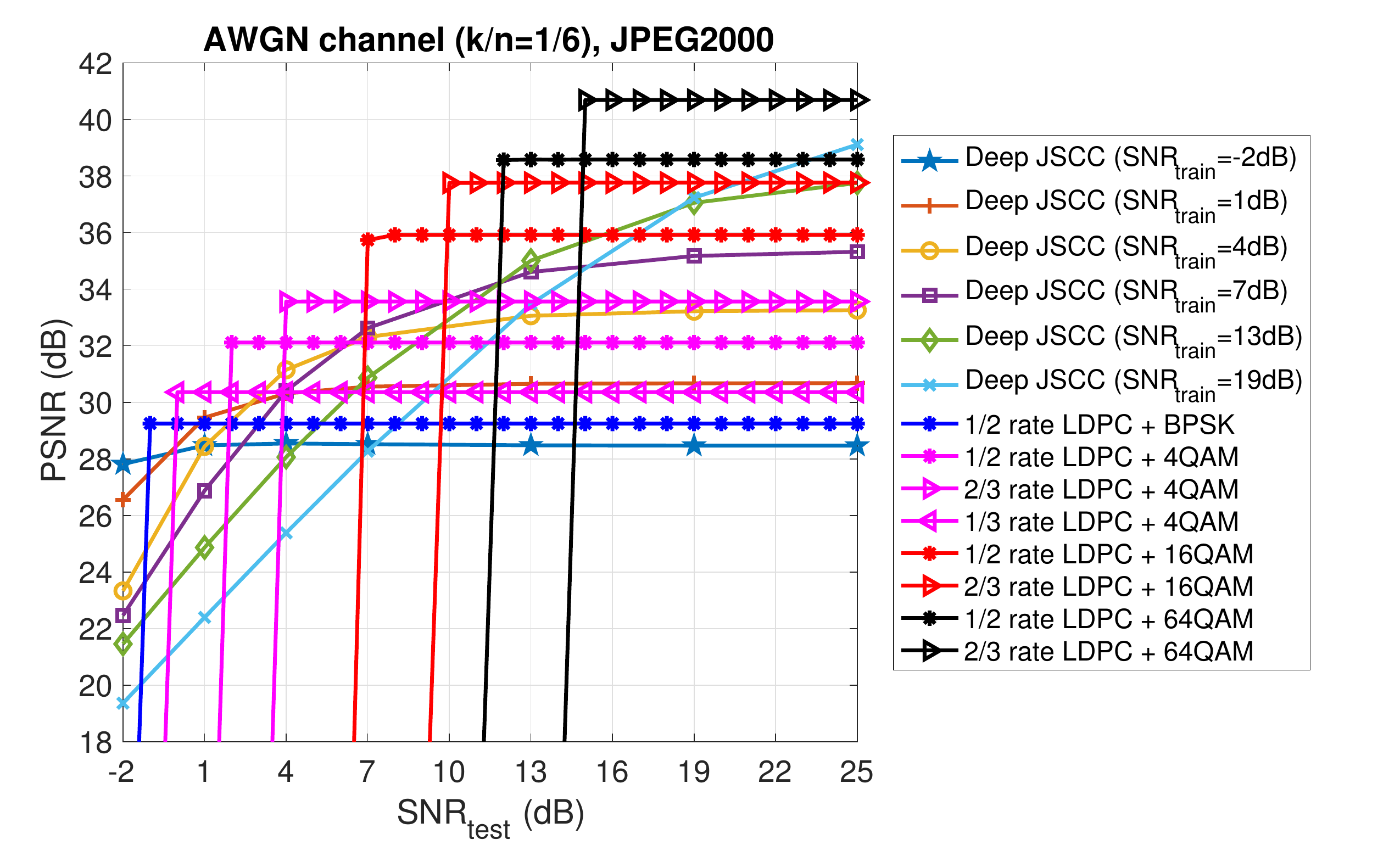}\label{fig:kodak_awgn_1over6_jpeg2000}}
	\end{center}
		\caption{Performance comparison of deep JSCC with baseline digital transmission schemes on the Kodak image dataset over AWGN channels with bandwidth compression ratio $k/n=1/6$. The digital schemes employ (a) JPEG and (b) JPEG2000 for image compression and various channel codes and modulation schemes.}
	\label{fig:kodak_awgn_1over6}
\end{figure}

Figs. \ref{fig:kodak_awgn_1over12}  and \ref{fig:kodak_awgn_1over6} show the performance of the proposed deep JSCC scheme and the digital transmission schemes in an AWGN channel as a function of the test SNR for bandwidth compression ratios $k/n = 1/12$ and $k/n = 1/6$, respectively. The results illustrate that our deep JSCC scheme significantly outperforms the baseline digital transmission schemes that use JPEG (the most widely used image compression algorithm) for low channel bandwidth and low SNR regimes, while it performs on par with the benchmark schemes for high bandwidth and high SNR values. Most importantly, our deep JSCC scheme does not suffer from the ``cliff effect'' observed in the digital transmission schemes. The inefficacy of the latter stems from the fact that, once the channel code and modulation scheme have been selected for a target SNR value, the number of bits available for compression is fixed and, thus, the quality of the reconstructed images does not improve with SNR. At the same time, when the channel quality drops below the target SNR value, the channel code is not able to deal with the increasing error rate, which leads to significant degradation in the quality of the reconstructed images. Contrarily to the digital transmission schemes, our deep JSCC scheme exhibits a graceful degradation of performance when the channel quality drops below the target SNR value, while the performance does not saturate immediately when the channel conditions improve beyond the target SNR.

When compared to schemes that use JPEG2000 for source compression, our JSCC algorithm outperforms the benchmark digital transmission schemes in AWGN channels only in very low SNR regimes and for low channel bandwidth. However, we believe that by using a deeper neural network architecture, and by employing more sophisticated activation and loss functions the performance of the deep JSCC algorithm can be further improved.

\begin{figure}[t]
	\begin{center}
		\subfloat[]{\includegraphics[width= 0.5\textwidth]{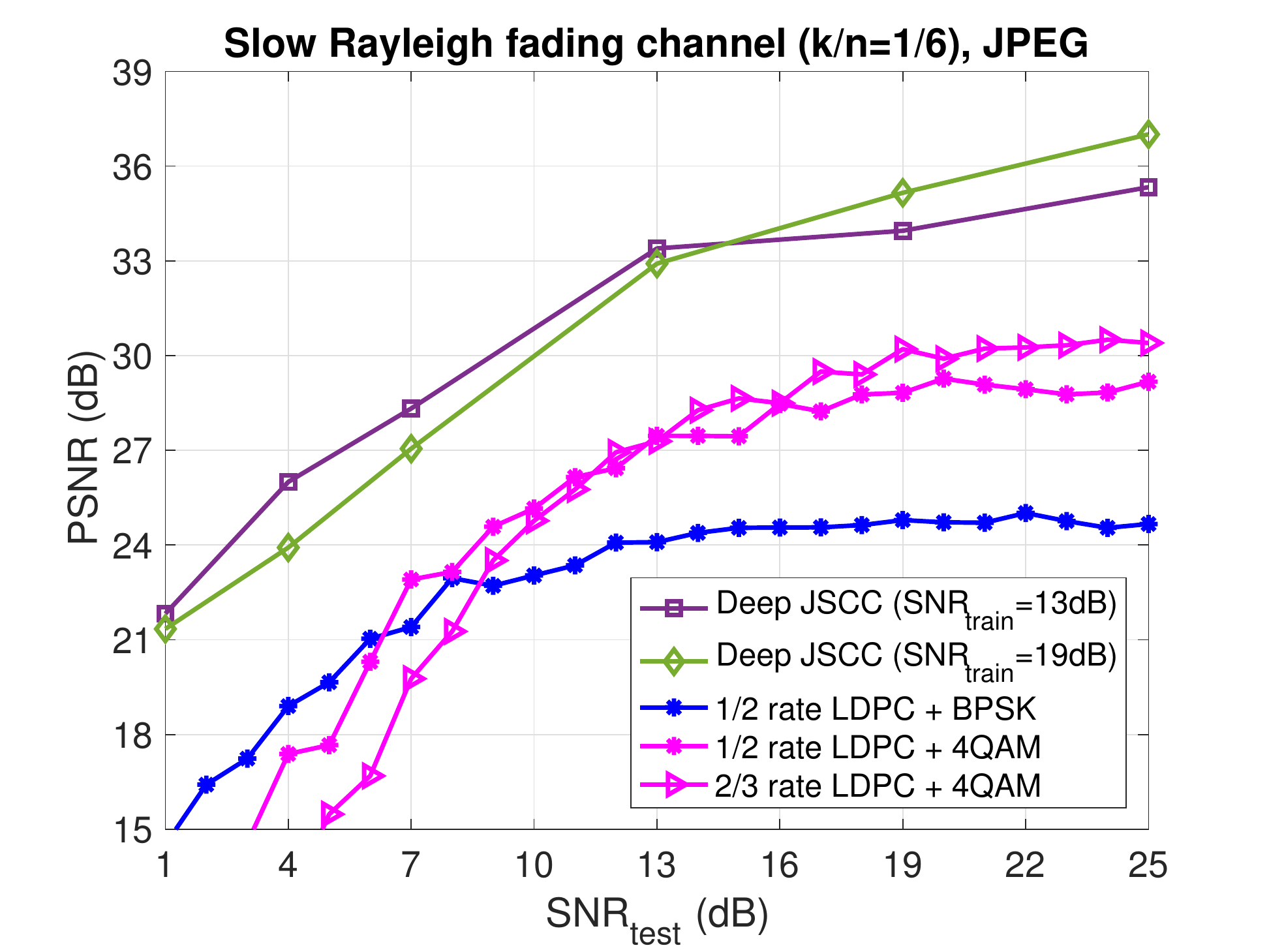}\label{fig:kodak_fading_1over6_jpeg}} \\
		\subfloat[]{\includegraphics[width=0.5\textwidth]{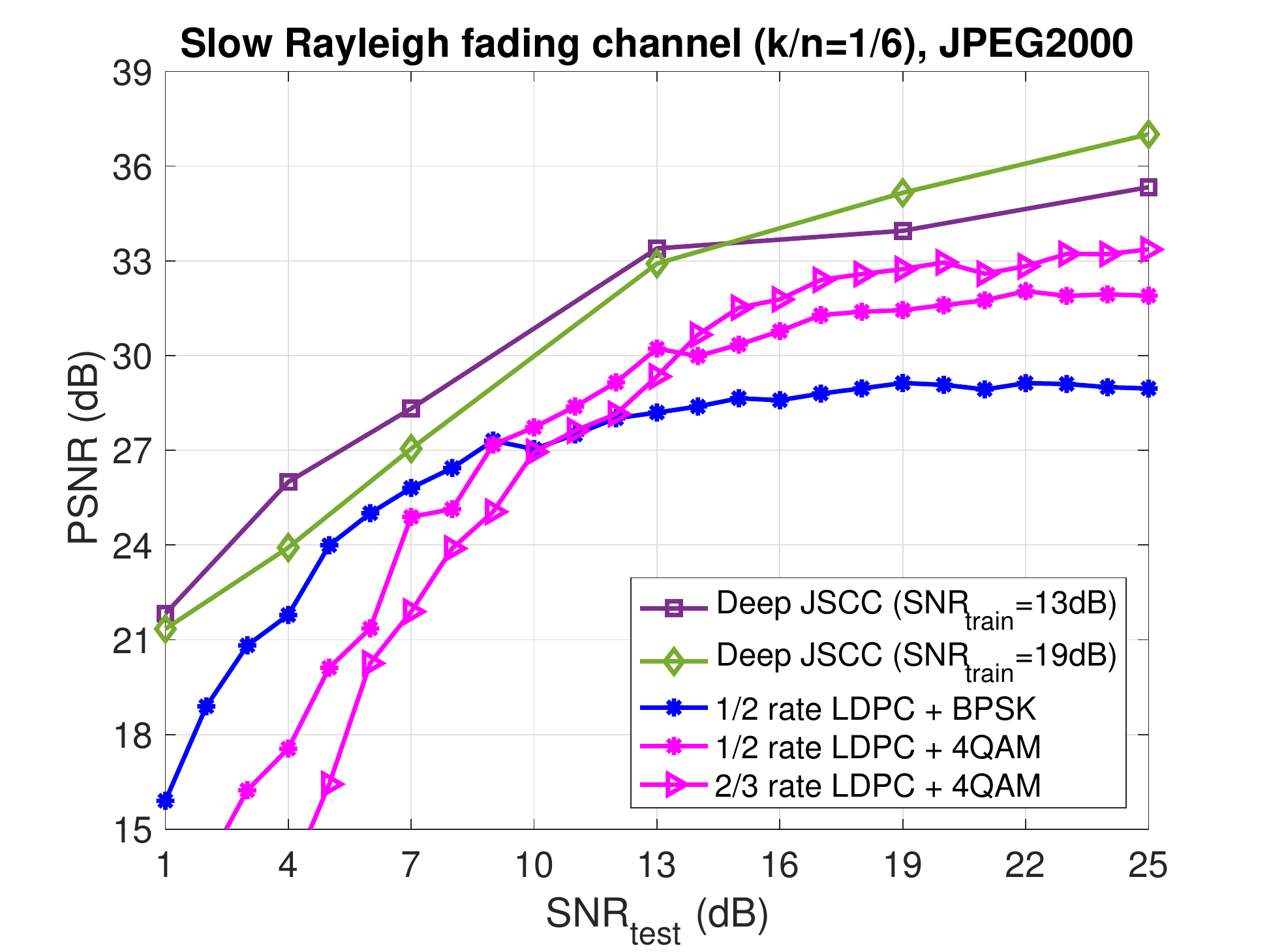}\label{fig:kodak_fading_1over6_jpeg2000}}
	\end{center}
		\caption{Performance comparison of deep JSCC with baseline digital transmission schemes on the Kodak image dataset over slow Rayleigh fading channels with bandwidth compression ratio $k/n=1/6$. The digital schemes employ (a) JPEG and (b) JPEG2000 for image compression and various channel codes and modulation schemes.}
	\label{fig:kodak_fading_1over6}
\end{figure}

We next evaluate the performance of our deep JSCC algorithm on the Kodak image dataset over time-varying channels. Fig. \ref{fig:kodak_fading_1over6} depicts the performance of deep JSCC and the benchmark digital transmission schemes in a slow Rayleigh fading channel for bandwidth compression ratio $k/n=1/6$.  We set the average channel gain to $H_c =1$ and vary the average SNR by varying the noise power $\sigma^2$. In these simulations, we assume that, in both the proposed scheme and the baseline digital transmission schemes, the phase shift introduced by the fading channel is known at the receiver, making the model equivalent to a real fading channel with double the bandwidth as only the channel gain changes randomly for each image transmission period. For the sake of readability, we only keep the best performing digital transmission schemes among all possible combinations of $1/2$, $2/3$ and $1/3$ rate LDCP codes and BSPK, 4-QAM, 16-QAM and 64-QAM modulation schemes. We can observe that due to the sensitivity of digital transmission schemes to the varying channel error rate as a result of varying channel SNR, the performance of the digital schemes that use separate source compression with JPEG/JPEG2000 followed by channel coding and modulation, is inferior to the performance of the proposed deep JSCC. While the digital transmission schemes perform well only in channel conditions for which they have been optimized, our deep JSCC scheme is more robust to channel quality fluctuations. Despite being trained for a specific average channel quality, deep JSCC is able to learn robust coded representations of the images that are resilient to fluctuations in the channel quality. The latter property is highly advantageous when transmitting over time-varying channels or to multiple receivers with different channel qualities.

\begin{figure*}
  \begin{center}
\begin{tabular}{cccc}
\textbf{Original} & \textbf{Deep JSCC} & \textbf{JPEG} & \textbf{JPEG2000}\\

\includegraphics[width=0.2\textwidth]{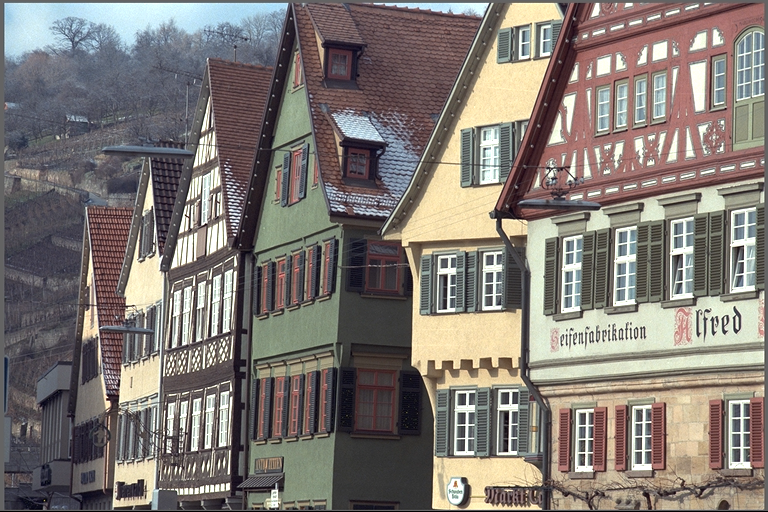}  &
\includegraphics[width=0.2\textwidth]{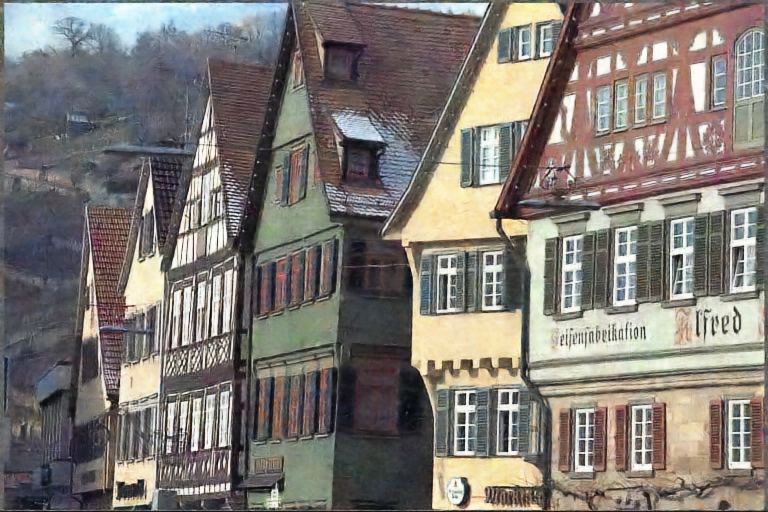}&
\includegraphics[width=0.2\textwidth]{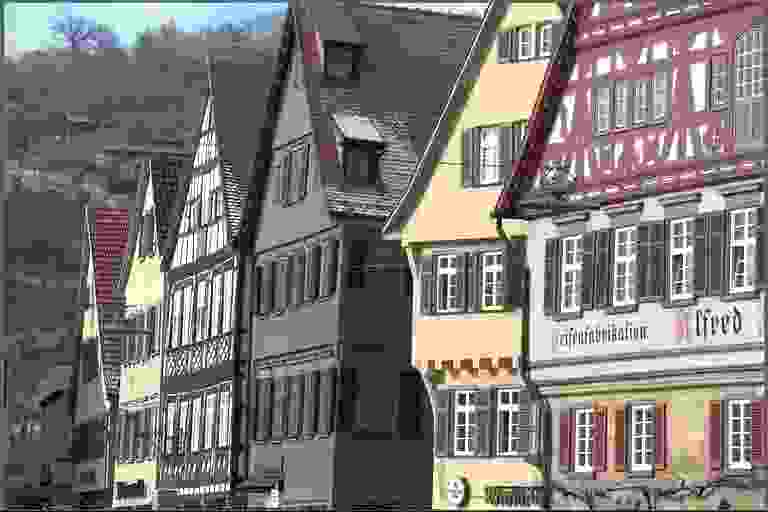}&
\includegraphics[width=0.2\textwidth]{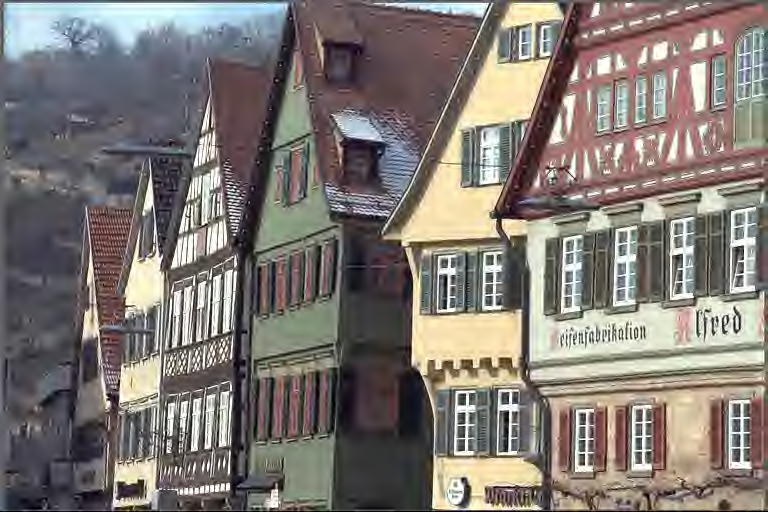}\\
PSNR/SSIM & 25.07dB/0.81 & 20.63dB/0.61 & 24.11dB/0.70 \\

\includegraphics[width=0.2\textwidth]{kodim08.png}  &
\includegraphics[width=0.2\textwidth]{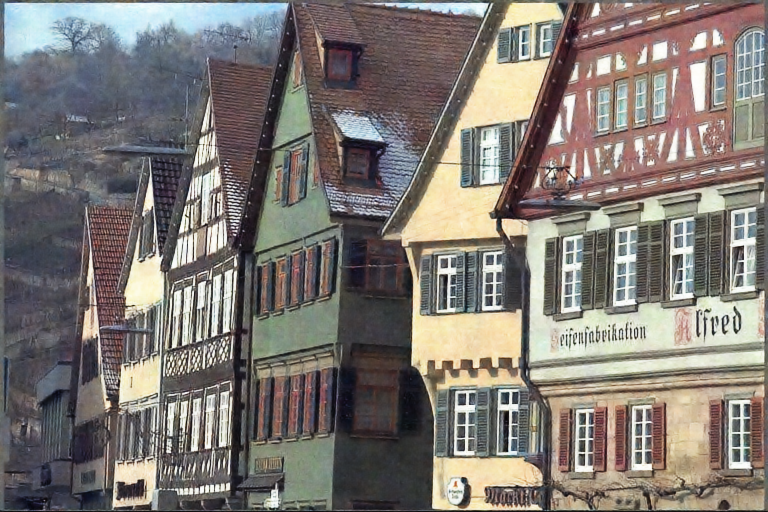}&
\includegraphics[width=0.2\textwidth]{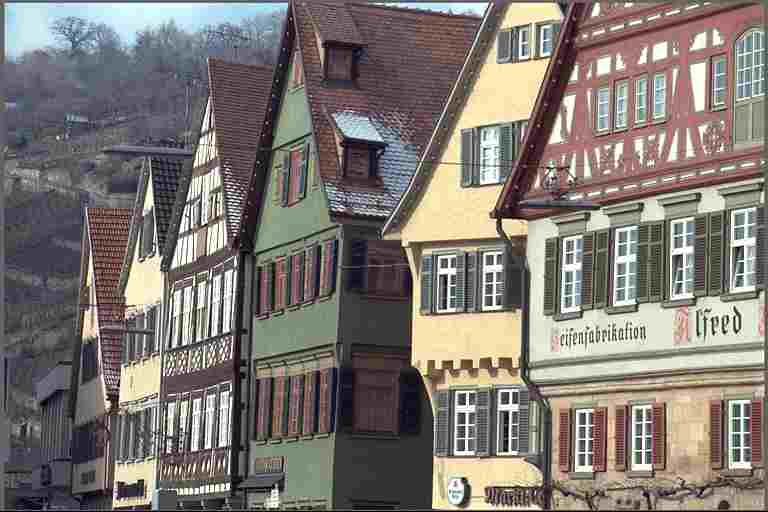}&
\includegraphics[width=0.2\textwidth]{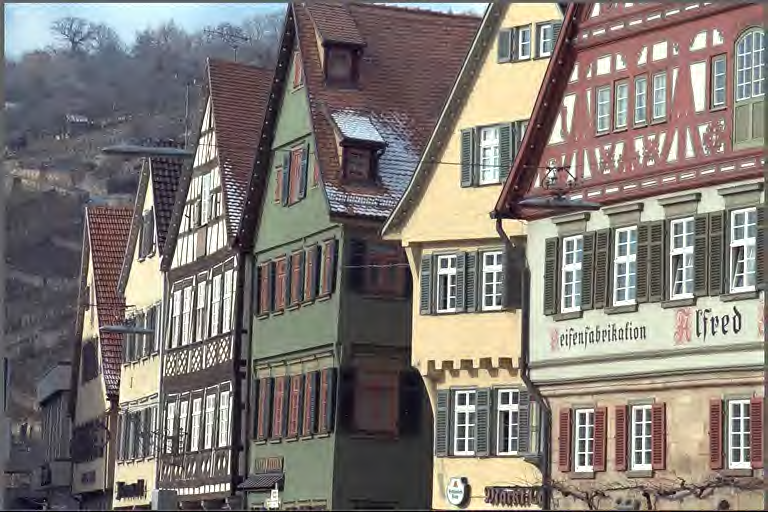}\\
PSNR/SSIM & 26.86dB/0.86 & 24.78dB/0.79 & 27.5dB/0.83 \\

\includegraphics[width=0.2\textwidth]{kodim08.png}&
\includegraphics[width=0.2\textwidth]{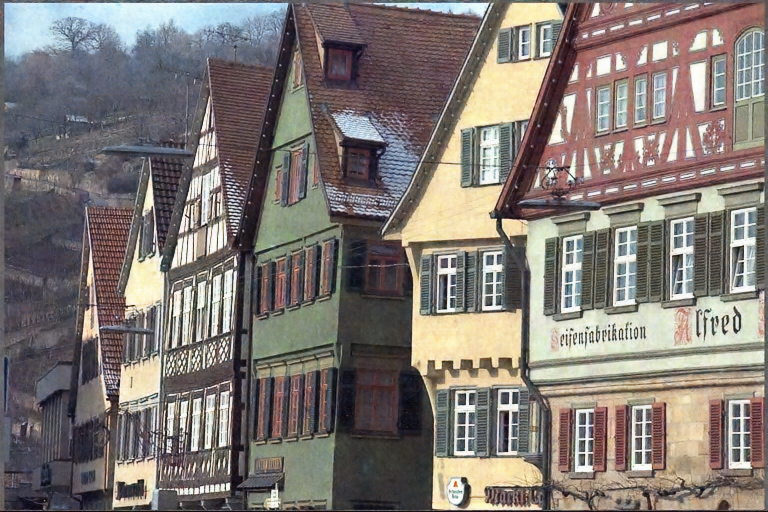}&
\includegraphics[width=0.2\textwidth]{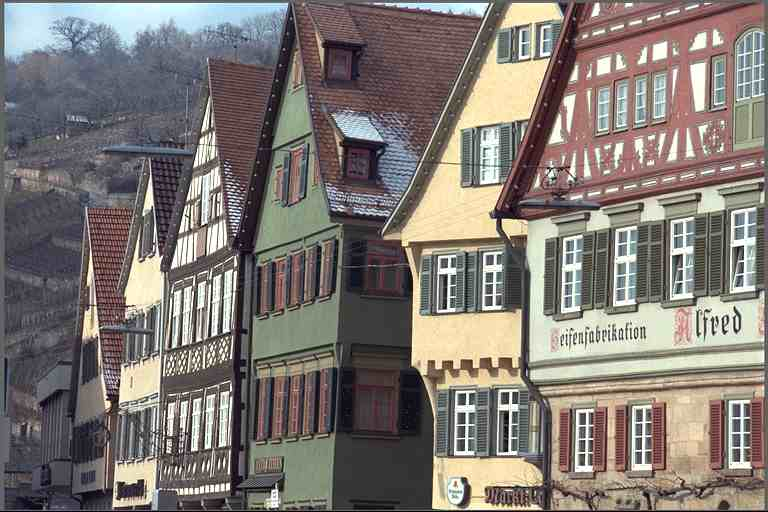}&
\includegraphics[width=0.2\textwidth]{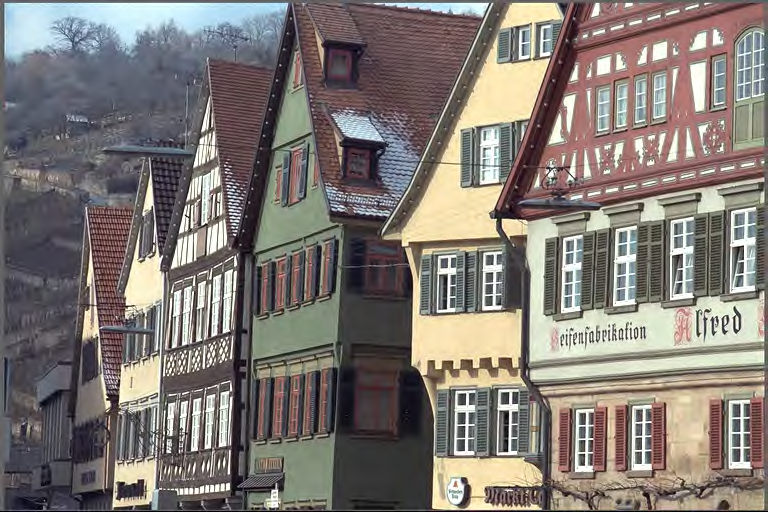}\\
PSNR/SSIM & 28.45dB/0.90 & 27.14dB/0.86 & 30.15dB/0.89 \\

\includegraphics[width=0.2\textwidth]{kodim08.png}&
\includegraphics[width=0.2\textwidth]{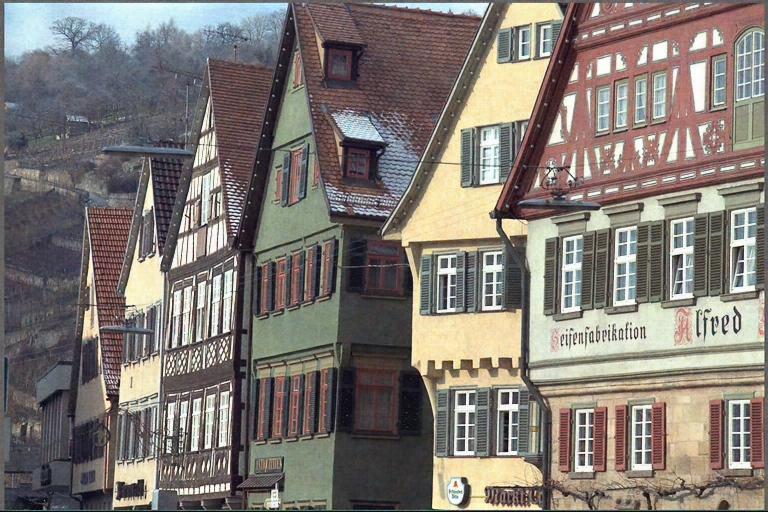}&
\includegraphics[width=0.2\textwidth]{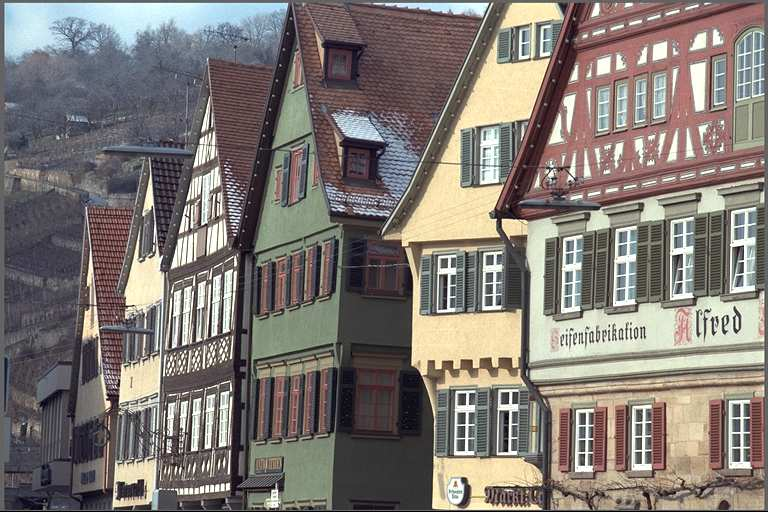}&
\includegraphics[width=0.2\textwidth]{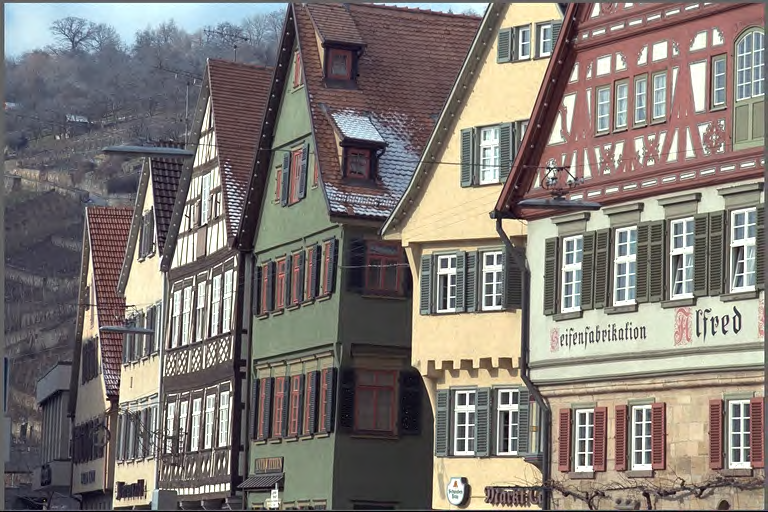} \\
PSNR/SSIM & 31.46dB/0.94 & 29.81dB/0.91 & 33.03dB/0.93 \\

\includegraphics[width=0.2\textwidth]{kodim08.png}&
\includegraphics[width=0.2\textwidth]{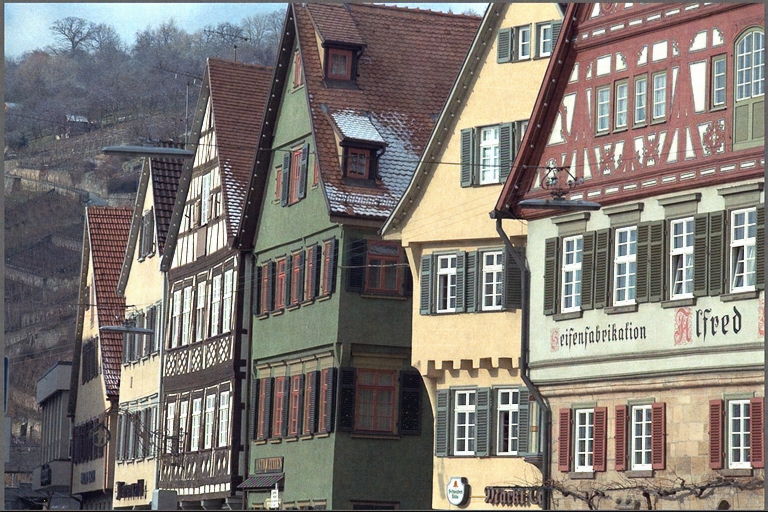}&
\includegraphics[width=0.2\textwidth]{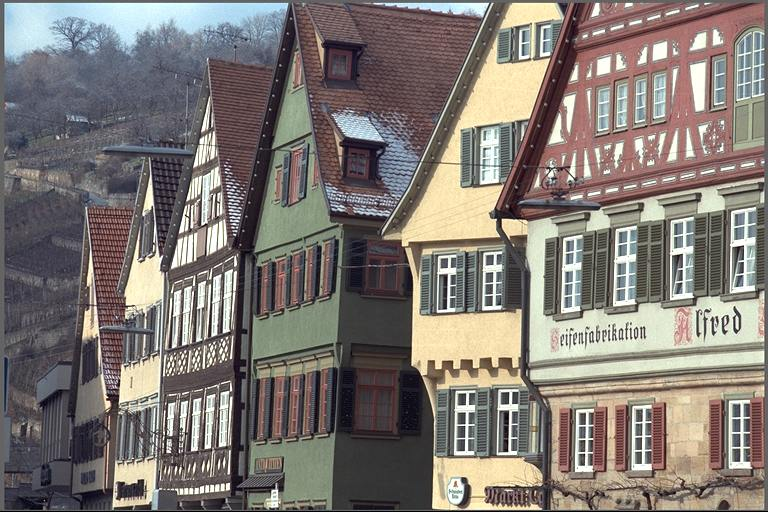}&
\includegraphics[width=0.2\textwidth]{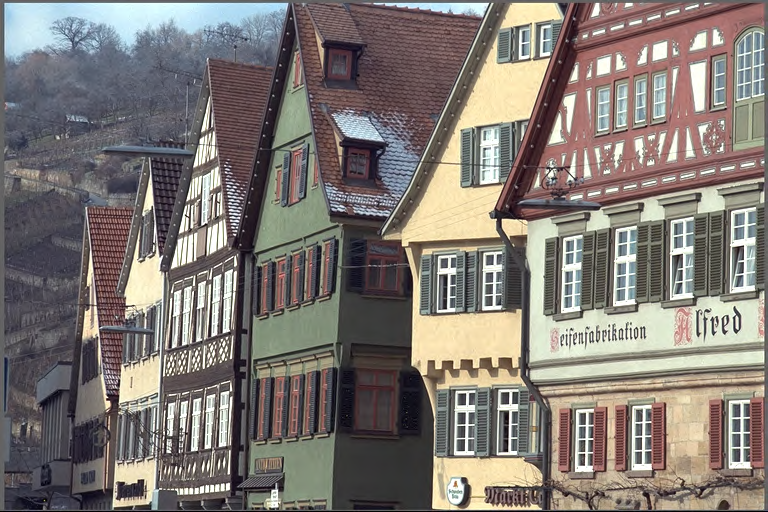} \\
PSNR/SSIM & 34.3dB/0.97 & 31.86dB/0.94 & 35.52dB/0.96

\end{tabular}
 \end{center}
    \caption{Examples of reconstructed images produced by the deep JSCC algorithm and the baseline digital schemes that use JPEG/JPEG2000 for image compression for AWGN channel and bandwidth compression ratio $k/n=1/6$. From top to bottom, the rows correspond to  SNR values of 1dB, 4dB, 7dB, 13dB and 19dB.}
    \label{fig:visual_kodak_awgn_1over6}
\end{figure*}

\begin{figure*}
  \begin{center}
\begin{tabular}{cccc}
\textbf{Original} & \textbf{Deep JSCC} & \textbf{JPEG} & \textbf{JPEG2000}\\
\includegraphics[width=0.2\textwidth]{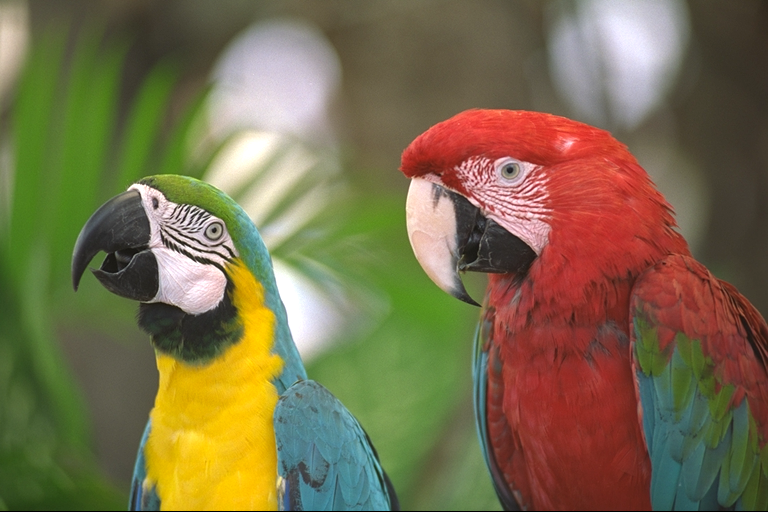}  &
\includegraphics[width=0.2\textwidth]{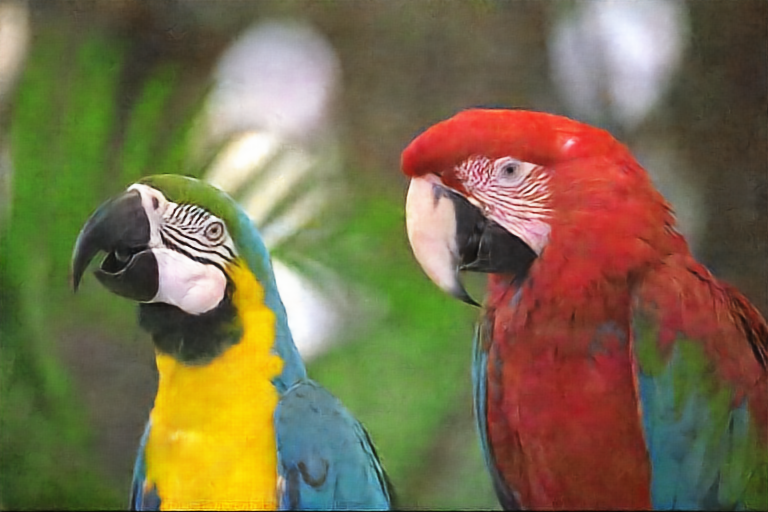}&
\includegraphics[width=0.2\textwidth]{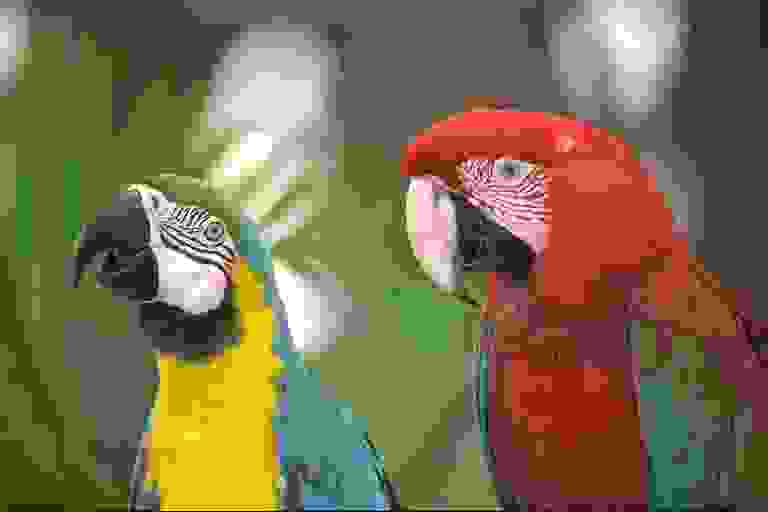}& N/A \\
 PSNR/SSIM & 30.69dB/0.87 & 22.68dB/0.67 &\\

\includegraphics[width=0.2\textwidth]{kodim23.png}  &
\includegraphics[width=0.2\textwidth]{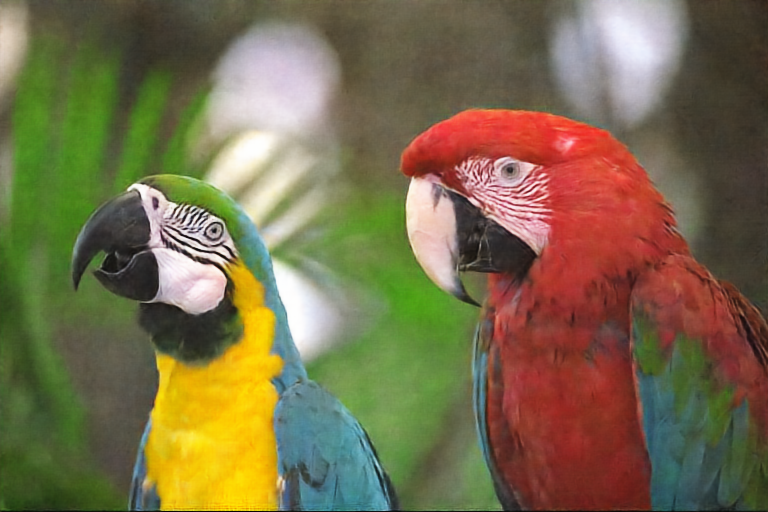}&
\includegraphics[width=0.2\textwidth]{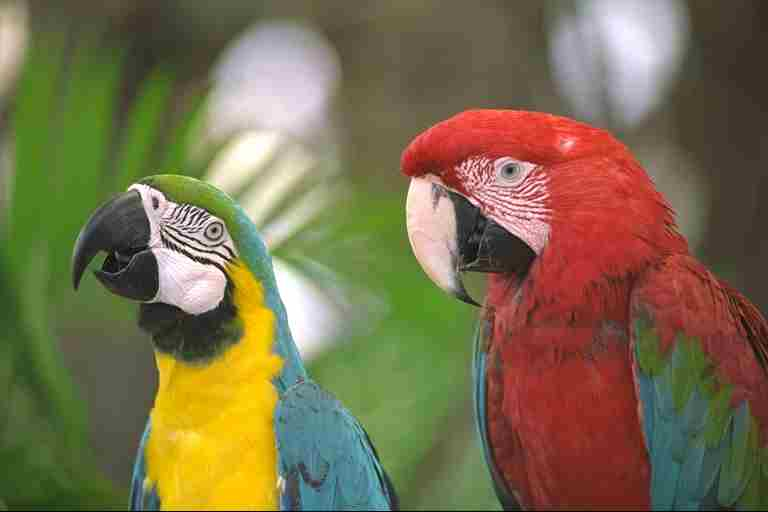}&
\includegraphics[width=0.2\textwidth]{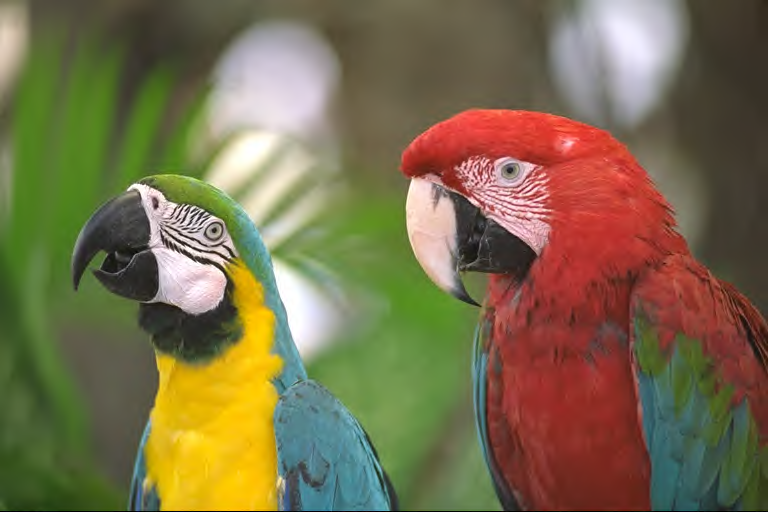}\\
 PSNR/SSIM & 31.92dB/0.89 & 31.65dB/0.86 & 36.40dB/0.92 \\

\includegraphics[width=0.2\textwidth]{kodim23.png}&
\includegraphics[width=0.2\textwidth]{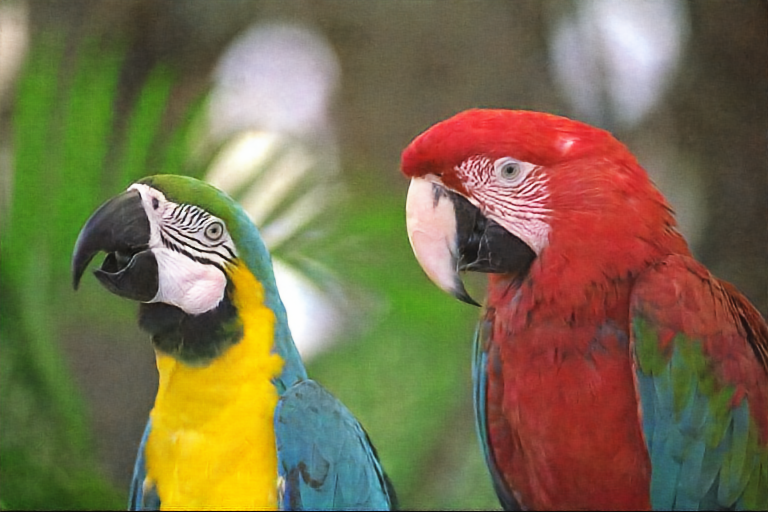}&
\includegraphics[width=0.2\textwidth]{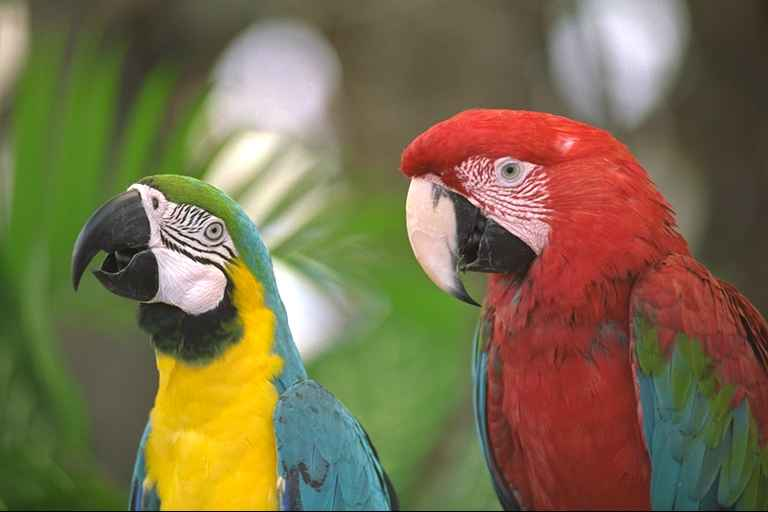}&
\includegraphics[width=0.2\textwidth]{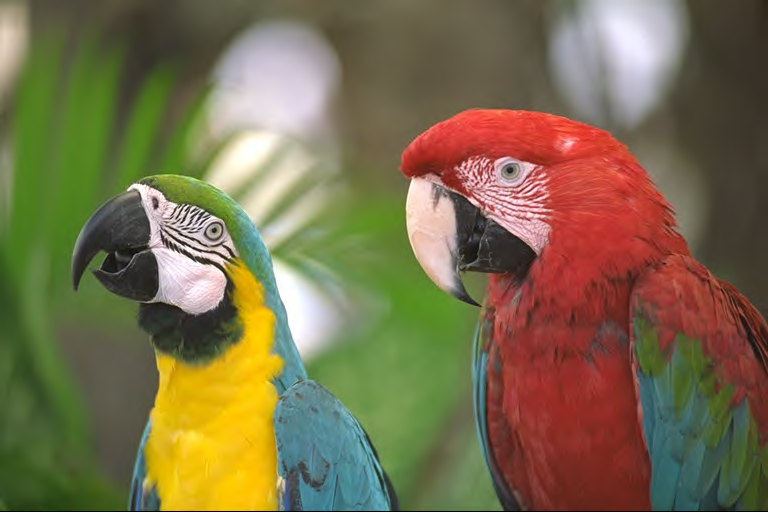}\\
 PSNR/SSIM & 32.90dB/0.90 & 34.36dB/0.91 & 38.46dB/0.94 \\

\includegraphics[width=0.2\textwidth]{kodim23.png}&
\includegraphics[width=0.2\textwidth]{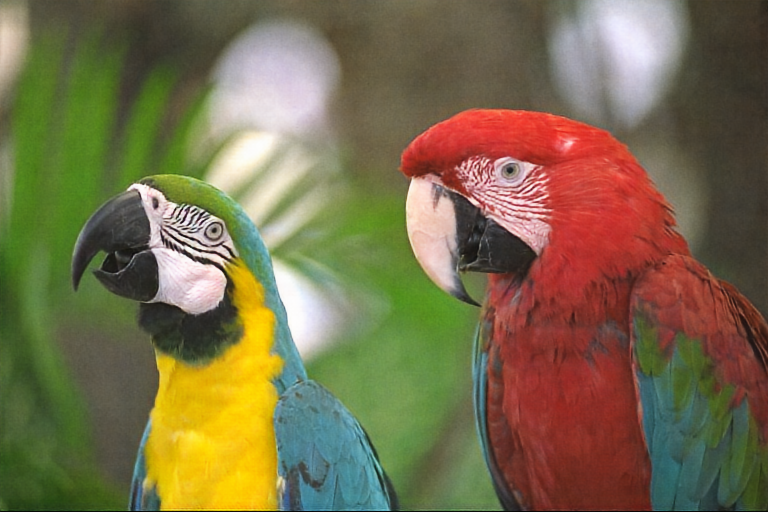}&
\includegraphics[width=0.2\textwidth]{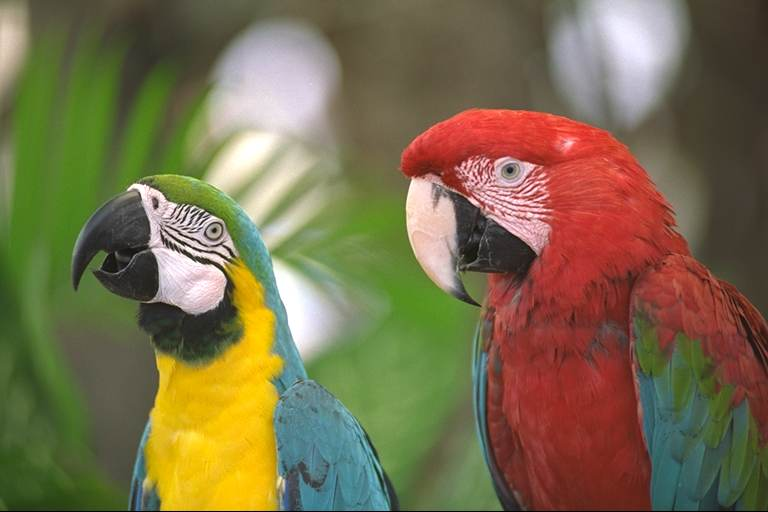}&
\includegraphics[width=0.2\textwidth]{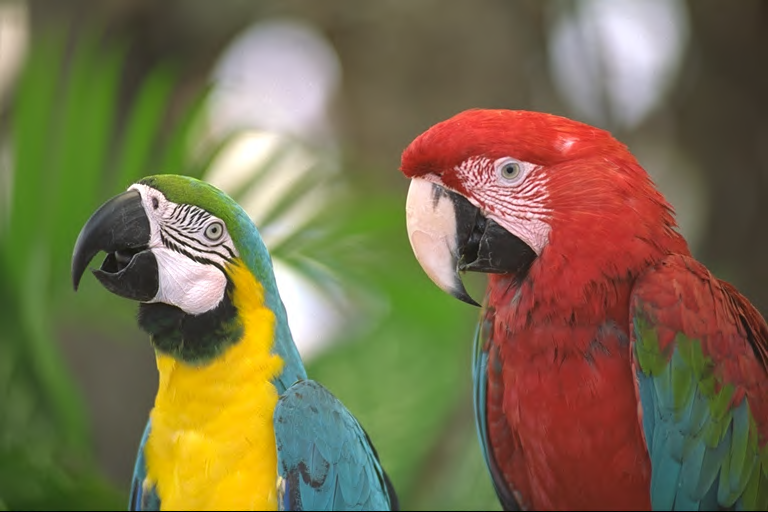} \\
 PSNR/SSIM & 35.34dB/0.93 & 36.45dB/0.93 & 40.5dB/0.96 \\

\includegraphics[width=0.2\textwidth]{kodim23.png}&
\includegraphics[width=0.2\textwidth]{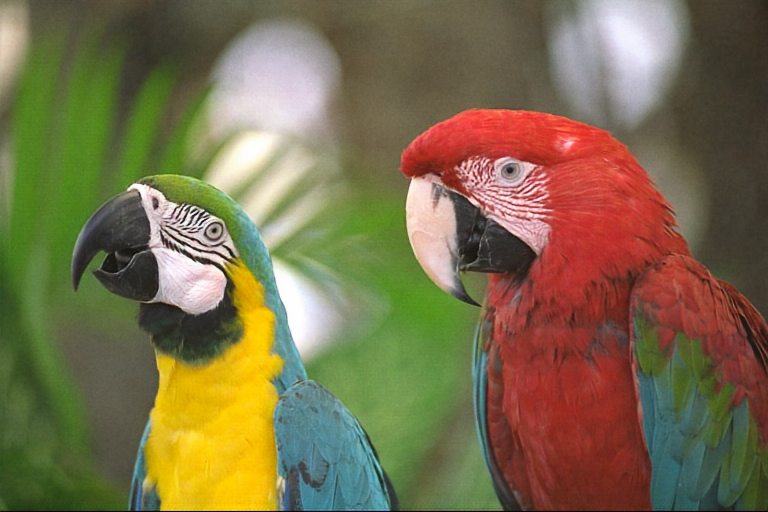}&
\includegraphics[width=0.2\textwidth]{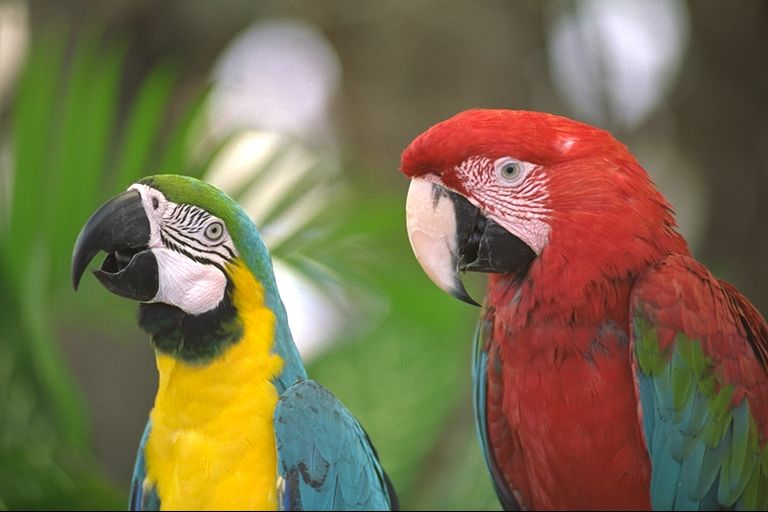}&
\includegraphics[width=0.2\textwidth]{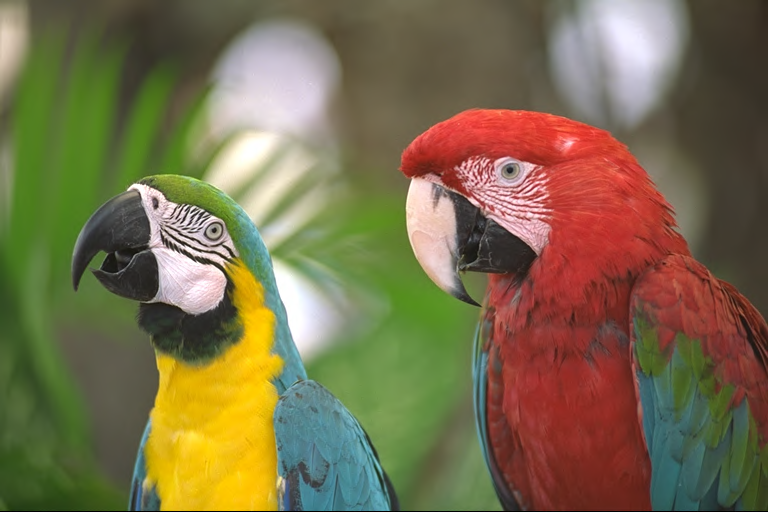} \\
 PSNR/SSIM & 36.83dB/0.94 & 37.79dB/0.95 & 41.96dB/0.96 \\

\end{tabular}
 \end{center}
    \caption{Examples of reconstructed images produced by the deep JSCC algorithm and the baseline digital schemes that use JPEG/JPEG2000 for image compression for AWGN channel and bandwidth compression ratio $k/n=1/12$. From top to bottom, the rows correspond to  SNR values of 1dB, 4dB, 7dB, 13dB and 19dB.}
    \label{fig:visual_kodak_awgn_1over12}
\end{figure*}

Finally, a visual comparison of the reconstructed images for the source and channel coding schemes under consideration in AWGN channels is presented in Figs. \ref{fig:visual_kodak_awgn_1over6} and \ref{fig:visual_kodak_awgn_1over12}. For the digital transmission schemes deploying JPEG/JPEG2000, the images are transmitted using the best-performing separate source and channel coding scheme for the target SNR value. Each row corresponds to a different channel SNR value starting from low SNR at the top (1dB) and progressing to high SNR (19dB) at the bottom. For each reconstruction, we report the PSNR and the SSIM \cite{Wang2004image} values. Fig. \ref{fig:visual_kodak_awgn_1over6}  illustrates an example where the deep JSCC outperforms the best performing digital scheme that deploys JPEG for source compression in terms of PSNR. More interestingly, although deep JSCC presents worse performance in terms of PSNR when compared to the separate scheme employing JPEG2000, its SSIM values are consistently higher, indicating superior perceived visual quality. Fig. \ref{fig:visual_kodak_awgn_1over12} shows an example where for high SNR values the digital transmission schemes outperform deep JSCC in the PSNR metric, but deep JSCC can still achieve comparable SSIM values when compared to the scheme using JPEG.  We can see that JPEG produces visible blocking artefacts, especially in channels with low SNR, which are not present in the images transmitted with deep JSCC. The noise introduced by deep JSCC appears to be smoother than the noise of JPEG thanks to the direct mapping of source values to soft channel input values. Note that the deep JSCC can also be trained with SSIM as the loss function, which can further improve its performance in terms of the SSIM metric.

\subsection{Computational complexity}
\label{sec:complexity}

In this section, we provide a brief discussion of the computational complexity of the proposed JSCC algorithm. Let us first consider the proposed encoder/decoder network. The most computationally costly operations in the encoder/decoder are the 2D convolutions/transpose convolutions, as they involve multiplications and additions. The computational cost of a single convolutional layer is  $ F \times F \times D \times K \times W \times H$ \cite{HowardARXIV2017}, where $F$ is the filter size, $K$ is the number of filters, $D$ is the number of input channels and $W \times H$ is the size of the feature map. The computational complexity of the encoder/decoder network is, thus,
$\mathcal{O} \Big{(}  I_W I_H \Big{)}$ where $I_W$ and $I_H$ are the input image width and height, respectively. This implies that the computational complexity of the proposed encoder/decoder is  linear in the number of pixels of the input image, as only the feature map width and height depend on the image dimensions, while all other factors are constant and independent of the image size. The JPEG encoding/decoding complexity is also linear in the number of pixels \cite{ChiouCEEC2017}, while LDPC codes have linear encoding/decoding times \cite{UrbankeMCT}. Thus, the computational complexity of a separate joint source and channel coding scheme, which employs JPEG for compression and LDPC codes for channel coding, is also linear in the size of the input image, i.e.,  $\mathcal{O} \Big{(}  I_W I_H \Big{)}$.

To complete our discussion of computational complexity, we have measured the average run time of the proposed algorithm on a Linux server with eight 2.10GHz Intel Xeon E5-2620V4 CPUs  and a Tesla K80 GPU. The measurements were performed on the Kodak color images with a resolution of $768 \times 512$ pixels. The average run time refers to the average time required to encode and decode one image using the proposed deep JSCC architecture. The average run time achieved by our GPU implementation is 18ms per image, while the average run time on CPU is 387ms. As a comparison, the average time required for the JPEG encoding and decoding of the above images, as reported in the literature, varies from 30ms \cite{RippelICML2017} to  390ms
\cite{ChengPCS2018}, while for the JPEG2000 algorithm the average encoding and decoding time on these images is even higher (e.g., 430-590ms \cite{RippelICML2017,ChengPCS2018}). This time must be further augmented by the time needed to encode/decode the compressed bitstream with a channel code. The above proves that our method is competitive with the baseline separate source and channel coding approaches not only in terms of quality, but also in terms of computational complexity.

\section{Conclusions and Future Work}\label{s:conclusions}

We have proposed a novel deep JSCC architecture for image transmission over wireless channels. In this architecture, the encoder maps the input image directly to channel inputs. The encoder and the decoder functions are modeled as complementary CNNs, and trained jointly on the dataset to minimize the average MSE of the reconstructed image. We have compared the performance of this deep JSCC scheme with conventional separation-based digital transmission schemes, which employ widely used image compression algorithms followed by capacity-achieving channel codes. We have shown through extensive numerical simulations that deep JSCC  outperforms separation-based schemes, especially for limited channel bandwidth and low SNR regimes. More significantly, deep JSCC is shown to provide a graceful degradation of the reconstruction quality with channel SNR. This observation is then used to benefit from the proposed scheme when communicating over a slow fading channel; deep JSCC performs reasonably well at all average SNR values, and outperforms the proposed separation-based transmission scheme at any channel bandwidth value.

In the case of DL-based JSCC, the encoder and decoder networks learn not only to communicate reliably over the channel (as in \cite{deep:PHY, Nachmani:JSTSP:18}), but also to compress the images efficiently. For a perfect channel with no noise, if the source bandwidth is greater than the channel bandwidth, i.e., $n>k$, the encoder-decoder NN pair is equivalent to an \textit{undercomplete autoencoder} \cite{GoodfellowDL2016}, which effectively learns the most salient features of the training dataset. However, in the case of a noisy channel, simply learning a good low-dimensional representation of the input is not sufficient. The network should also learn to map the salient features to nearby representations so that similar images can be reconstructed despite the presence of noise. We also note that, the resilience to channel noise acts as a sort of a regularizer for the autoencoder. For example, when there is no channel noise, if the channel bandwidth is larger than the source bandwidth, i.e., $n<k$, we obtain an \textit{overcomplete autoencoder}, which can simply learn to replicate the image. However, when there is channel noise, even an overcomplete autoencoder learns a non-trivial mapping that is resilient to channel noise, similarly to denoising autoencoders.

The next step in improving the performance of the deep JSCC scheme is to exploit more advanced NN architectures in the autoencoder that have been shown to improve the compression performance \cite{Balle:ICLR:17,Johnston:CVPR:18}. We will also explore the performance of the system for non-Gaussian channels as well as for channels with memory, for which we do not have capacity-approaching channel codes. We expect that the benefits of the proposed NN-based JSCC scheme will be more evident in these non-ideal settings.

\bibliographystyle{IEEEtran.bst}

 \end{document}